\documentclass[iop]{emulateapj}

\usepackage{color}
\usepackage{ulem}
\usepackage{dcolumn}
\usepackage[flushleft]{threeparttable}

\renewcommand{\prl}{PhRvL}
\renewcommand{\gca}{GeCoA}
\renewcommand{\nat}{Natur}
\renewcommand{\prc}{PhRvC}
\renewcommand{\apjl}{ApJL}

 \shorttitle{nucleosynthesis in neutrino-driven wind in HNe}
 \shortauthors{S.Fujibayashi, T.Yoshida and Y.Sekiguchi}

\begin{document}

\title{Nucleosynthesis in neutrino-driven winds in hypernovae}

\author{
Sho~Fujibayashi
}
\affil{
Department of Physics, Graduate School of Science, Kyoto University, Kyoto 606-8502, Japan
}
\email{
fujibayashi@tap.scphys.kyoto-u.ac.jp
}

\author{
Takashi~Yoshida
}
\affil{
Yukawa Institute for Theoretical Physics, Kyoto University, Kyoto 606-8502, Japan
}

\author{
Yuichiro~Sekiguchi
}
\affil{
Department of Physics, Toho University, Funabashi, Chiba 274-8510, Japan
}


\begin{abstract}
We investigate the nucleosynthesis in the neutrino-driven winds blown off from a $3M_\odot$ massive proto-neutron star (mPNS) temporarily formed during the collapse of a $100M_\odot$ star.
Such mPNSs would be formed in hypernovae.
We construct steady and spherically symmetric wind solutions. We set large neutrino luminosities of $\sim 10^{53}\ {\rm erg\ s^{-1}}$ and average energies of electron neutrinos and antineutrinos in the ranges of $\epsilon_{\nu_e}=9-16\ {\rm MeV}$ and $\epsilon_{\bar{\nu}_e}=11-18\ {\rm MeV}$ based on a recent numerical relativity simulation.
The wind solutions indicate much shorter temperature-decrease timescale than that of the winds from ordinary PNSs and, depending on $\epsilon_\nu$, the winds can be both neutron-rich and proton-rich.
In the neutron-rich wind, the $r$-process occurs and the abundance distribution of a fiducial wind model of the mPNS gives an approximate agreement with the abundance pattern of metal-poor weak $r$ star HD~122563, although the third-peak elements are produced only when the $\bar{\nu}_e$ energy is much larger than the $\nu_e$ energy.
In the proton-rich wind, the strong $\nu p$-process occurs and $A>100$ nuclides are synthesized. The synthesized nuclei can be neutron-rich in some cases because the large neutrino luminosity of the mPNS supplies a sufficient amount of neutrons.
\end{abstract}

\section{Introduction}

The neutrino-driven wind from nascent neutron stars in core-collapse 
supernovae (SNe) is a possible site of heavy-element synthesis.
A number of studies about the $r$-process, which produces about a half of 
elements heavier than iron, in the neutrino-driven wind have been made in 
the past decades \citep[e.g.,][]{1996ApJ...471..331Q, 1997ApJ...482..951H, 
2000ApJ...533..424O, 2001ApJ...562..887T, 2001ApJ...554..578W, 2010ApJ...722..954R}.
However, it was revealed that the strong $r$-process that makes all nuclei up 
to the third-peak elements ($A\approx 195$) hardly occurs in the standard condition of the neutrino-driven wind.
The most critical issue toward the successful $r$-process is that the electron fraction $Y_e$ of the neutrino-driven wind is larger than 0.5, i.e., the wind is not neutron-rich as recent numerical simulations of proto-neutron star (PNS) evolution suggested \citep[e.g.,][]{2010A&A...517A..80F,2010PhRvL.104y1101H}.
Electron fraction in the neutrino-driven wind still has uncertainties due to the input physics such as mean field effects on neutrino opacities \citep[e.g.,][]{2012PhRvC..86f5806H,2012PhRvC..86f5803R}, so the $r$-process might occur in the wind.
Nevertheless, recent studies have indicated that the $r$-process in such a wind is too weak to produce the third-peak elements \citep[e.g.,][]{2013ApJ...770L..22W}.

Instead, the proton-rich wind is a promising site of the $\nu p$-process, which is the nucleosynthesis process found in the 2000's.
Recent studies show that some light proton-rich nuclei can be made in hot bubbles and neutrino-driven winds through the $\nu p$-process \citep{2006PhRvL..96n2502F,2006ApJ...644.1028P, 2006ApJ...647.1323W, 2011ApJ...729...46W}.

In the 1990's, hypernovae (HNe), more energetic explosive events than ordinary SNe, were discovered \citep{1998Natur.395..670G,1998Natur.395..672I}.
Although the explosion mechanism of HNe is yet unclear, there are some observational hints about their progenitors.
First, progenitors of HNe are thought to be massive stars, with the initial mass larger than $\sim 25M_\odot$ \citep[e.g., HN branch in][]{2006NuPhA.777..424N}, which is larger than the masses of the progenitors of ordinary SNe.
Second, at least some of HNe show the features of bipolar explosion \citep{2006ApJ...640..854M,2006ApJ...645.1331M}, which indicates that the rotation of the progenitor may play a role in the explosion.
Based on these studies, we assume here that HNe are bipolar explosions associated with the death of very massive stars.

After the explosion, the neutrino-driven wind may be launched toward the direction of the bipolar explosion if a PNS is formed at least for a while.
Indeed, according to recent numerical simulations of the collapse of rotating massive stellar cores \citep{2012PTEP.2012aA304S,2013ApJ...779L..18C}, a massive proto-neutron star (mPNS) is temporarily formed after the collapse \citep[see also][for spherical collapse]{2007ApJ...667..382S,2008ApJ...688.1176S}.
The mPNS is supported against the collapse by its thermal pressure and rotation and eventually collapses into a black hole.
The lifetime of the mPNS is a few seconds even in the case of  the collapse of a $100M_\odot$ star \citep{2012PTEP.2012aA304S} for an equation of state (EOS) with which the maximum mass of the neutron star is larger than $2M_\odot$.
For a less massive progenitor, the lifetime would be much longer.

Based on the above picture for HNe and the collapse of a massive star, we consider neutrino-driven wind from the mPNS in the HNe.
We propose the {\it HN wind model} as follows:
We assume that an explosion occurs toward the polar region when a massive star collapses.
After the explosion, the neutrino-driven wind is launched from the mPNS surface toward the polar region.
For the mPNS model, we adopt the result in \cite{2012PTEP.2012aA304S}.

There are two main differences between HNe and ordinary SNe.
One of them is the (temporal) formation of more massive PNSs in HNe for the consequence of the collapse of the massive star.
The other is the higher neutrino luminosity in HNe, which is expected to be provided by the stellar-gas accretion to the mPNS along the directions that are not the direction of the bipolar explosion (e.g., equator in rotational collapse).
In addition to these, as some models of HNe \citep[e.g.,][]{2007ApJ...664..416B} suggest, the rotation and the magnetic field of the PNS could be important for HN explosion.
We will discuss these effects in Section \ref{sec:discussion}.

In this work, based on this scenario, we investigate the nucleosynthesis in the HN wind.
It is known that the entropy of the neutrino-driven wind strongly depends on the mass of a PNS because of the general relativistic effects \citep{1997ApJ...486L.111C, 2000ApJ...533..424O}.
Generally speaking, a high-entropy wind results in more efficient heavy-element synthesis because less seed elements are produced through triple-$\alpha$ and ${}^{4}{\rm He}(\alpha n,\gamma){}^{9}{\rm Be}(\alpha,n){}^{12}{\rm C}$ reactions \citep{1997ApJ...482..951H}.
When the abundance of the seed elements is small, the neutron-to-seed ratio becomes large and thus further heavy-element synthesis proceeds.
The expansion timescale has a crucial role on heavy-element synthesis.
A shorter expansion timescale also causes less effective seed element production.
According to the numerical simulation of the collapse of the very massive star and the mPNS formation \citep{2012PTEP.2012aA304S}, the neutrino luminosity of the mPNS is much higher than that of an ordinary PNS as mentioned above.
In this case, the expansion velocity of the neutrino-driven wind of the mPNS is also expected to be faster than that of the PNS, so the timescale of seed element production is also shorter than that of the PNS.
Thus, the neutrino-driven wind from an mPNS, although the lifetime of the mPNS may be short, is expected to be an interesting site of heavy-element synthesis.

In this work, we investigate possible nucleosynthesis processes to produce heavy elements in the neutrino-driven winds from mPNSs.
Our model of the neutrino-driven winds and the wind parameters are described in Section \ref{sec:neutrino-driven-wind-model}.
A brief description of the nucleosynthesis calculation is given in Section \ref{sec:nucleosynthesis-calculation}.
The results of the nucleosynthesis calculation and details of the nucleosynthesis processes are given in Section \ref{sec:results}.
Finally, discussions and conclusions are given in Sections \ref{sec:discussion} and \ref{sec:conclusions}.
Throughout this paper, we use the unit in which the gravitational constant $G$, the speed of light $c$, and Planck constant $\hbar$ are taken to be unity.

\section{Neutrino-driven wind model}
\label{sec:neutrino-driven-wind-model}
\subsection{Basic Equations}

The basic equations of the fluid are the equations of momentum conservation, baryon number conservation, and energy conservation.
We ignore the self-gravity of the fluid and the change of the mass of the neutron star due to the mass ejection through the wind and the gas accretion through the equatorial plane.
For simplicity, we consider a steady and spherically symmetric flow, so that the gravitational fields can be treated as those in the Schwarzschild spacetime.
We do not consider the effects of the rotation and the magnetic field of the PNS, which will be discussed in Section \ref{sec:discussion}.
Then, the equations can be written in the Schwarzschild coordinate as
\begin{eqnarray}
\dot{M} &=& 4\pi r^2 \rho u, \label{eq:wind-mdot}\\
\frac{du}{d\tau} &=& u \frac{du}{dr} = -\frac{1+u^2 -2M/r}{\rho(1+\epsilon)+P}\frac{dP}{dr} - \frac{M}{r^2},\label{eq:wind-mom}\\
T\frac{ds}{d\tau} &=& u\left(\frac{d\epsilon}{dr} - \frac{P}{\rho ^2}\frac{d\rho}{dr} \right) = \dot{q}, \label{eq:wind-qdot}
\end{eqnarray}
where $r$ is the radial coordinate, $M$ is the mass of the PNS, $\dot{M}$ is the mass ejection rate, which is constant in the steady wind, $u=dr/d\tau$ is the radial component of the four-velocity of the fluid element, $\tau$ is the proper time of the fluid element, $\rho$ is the rest-mass density, $P$ is the pressure, $T$ is the temperature, $\epsilon$ is the specific internal energy, $s$ is the entropy per unit mass, and $\dot{q}$ is the net heating rate per unit time per unit mass.
These equations are the same as those in \cite{2000ApJ...533..424O} and \cite{2001ApJ...554..578W}.
In addition to these equations, we solve the evolution equation for the electron fraction $Y_e$, which is fixed in their work.
$Y_e$ changes owing to the interactions between nucleons and neutrinos:
\begin{eqnarray}
\frac{dY_e}{d\tau} &=& u\frac{dY_e}{dr}\nonumber\\
&=& -\ (\lambda_{\bar{\nu}p}+\lambda_{\nu n}+\lambda_{\rm ec}+\lambda_{\rm pc})Y_e + (\lambda_{\nu n}+\lambda_{\rm pc}),\label{eq:wind-Ye}
\end{eqnarray}
where $\lambda_{\bar{\nu}p},\lambda_{\nu n},\lambda_{\rm ec},$ and $\lambda_{\rm pc}$ are the rates for the electron (anti)neutrino absorption reaction by free protons (neutrons) and electron (positron) capture reaction by free protons (neutrons), respectively. These reactions are written below:
\begin{eqnarray}
\bar{\nu}_e + p &&\rightleftarrows n + e^+, \label{reac:nup}\\
\nu_e + n &&\rightleftarrows p + e^-.\label{reac:nun}
\end{eqnarray}
The details of the calculation of these rates will be given in Section \ref{sec:heating-and-cooling-rates and-reaction-rates-due-to-weak-interactions}.

The net heating rate $\dot{q}$ appeared in Equation (\ref{eq:wind-qdot}) is composed of the terms of heating and cooling involving five processes and can be decomposed as \citep[e.g.,][]{1996ApJ...471..331Q}
\begin{eqnarray}
\dot{q}=\dot{q}_{\nu N}-\dot{q}_{eN}+\dot{q}_{e\nu}-\dot{q}_{e^+e^-}+\dot{q}_{\nu\bar{\nu}},
\end{eqnarray}
where $\dot{q}_{\nu N}$, $\dot{q}_{e\nu}$, and $\dot{q}_{\nu\bar{\nu}}$ are the heating rates due to neutrino absorption on free nucleons, neutrino-electron scattering, and neutrino-antineutrino pair annihilation into electron-positron pairs, and $\dot{q}_{eN}$ and $\dot{q}_{e^+e^-}$ are the cooling rates due to electron and positron capture by free nucleons and electron-positron pair annihilation into neutrino-antineutrino pairs. The details will be given in Section \ref{sec:heating-and-cooling-rates and-reaction-rates-due-to-weak-interactions}.

We rewrite Equations (\ref{eq:wind-mdot})-(\ref{eq:wind-Ye}) to obtain
\begin{eqnarray}
\frac{\partial \epsilon}{\partial T}(u^2-Ac_s^2)\,\frac{du}{dr} &=&u\frac{\partial \epsilon}{\partial T} \left(\frac{2A}{r}\frac{\partial P}{\partial \rho} - \frac{M}{r^2}\right)\nonumber \\
&&-\ \frac{A}{\rho}\frac{\partial P}{\partial T}\left[\dot{q} + \frac{2u}{r}\left(\rho\frac{\partial \epsilon}{\partial \rho}-\frac{P}{\rho}\right)\right] \nonumber \\
&&+\ \frac{Au}{\rho}\frac{dY_e}{dr}\left(\frac{\partial P}{\partial T}\frac{\partial \epsilon}{\partial Y_e} - \frac{\partial P}{\partial Y_e} \frac{\partial \epsilon}{\partial T}\right),\hspace{5mm}\label{eq:u-evol}\\
\frac{\partial \epsilon}{\partial T}(u^2-Ac_s^2)\,\frac{dT}{dr}&=&\dot{q}\left(u-\frac{A}{u}\frac{\partial P}{\partial \rho}\right) \nonumber \\
&&+\ \left(\rho\frac{\partial \epsilon}{\partial \rho}-\frac{P}{\rho}\right)\left(\frac{2u^2}{r}-\frac{M}{r^2}\right)\nonumber\\
&&-\ \frac{dY_e}{dr}\biggl[\frac{\partial \epsilon}{\partial Y_e}\left(u^2-A\frac{\partial P}{\partial \rho}\right)\nonumber\\
&&+\ \frac{A}{\rho}\frac{\partial P}{\partial Y_e}\left(\rho\frac{\partial \epsilon}{\partial \rho}-\frac{P}{\rho}\right)\biggr], \label{eq:T-evol}\\
u\frac{dY_e}{dr} &=& -\ (\lambda_{\bar{\nu}p}+\lambda_{\nu n}+\lambda_{\rm ec}+\lambda_{\rm pc})Y_e\nonumber\\
&&+\ (\lambda_{\nu n}+\lambda_{\rm pc}),\label{eq:Ye-evol}
\end{eqnarray}
where $A=A(r)$ is defined as
\begin{eqnarray}
A(r)=\frac{1+u^2 -2M/r}{1+\epsilon+P/\rho}.
\end{eqnarray}
Here $c_s$ is the sound velocity given by
\begin{eqnarray}
\hspace{-5mm}c_s^2 = \biggl(\frac{\partial P}{\partial \rho}\biggr)_{\hspace{-1mm}s} = \frac{\partial P}{\partial \rho} +\frac{1}{\rho}\frac{\partial P}{\partial T} \left(\frac{\partial \epsilon}{\partial T}\right)^{\hspace{-1mm}-1}\hspace{-1mm}\left(\frac{P}{\rho} - \rho \frac{\partial \epsilon}{\partial \rho}\right),
\end{eqnarray}
where the thermodynamic quantities $\epsilon$ and $P$ are thought to be the functions of $\rho, T$ and $Y_e$.

To close the system of equations, we adopt an EOS by \cite{2000ApJS..126..501T}, which includes the pressure of nucleons, electrons, positrons, and photons.

We calculate the proper time of the fluid element from the definition of the four-velocity as
\begin{eqnarray}
\tau(r) = \int^{r} \frac{dr'}{u(r')}.
\end{eqnarray}

We construct the wind solution by solving Equations (\ref{eq:u-evol})-(\ref{eq:Ye-evol}) from the gain radius $R_{\rm gain}$, where the net heating rate becomes zero ($\dot{q}=0$).
For simplicity, we assume that the radius of the neutrinosphere $R_\nu$ is equal to the gain radius $R_{\rm gain}$.
This assumption is justified by the fact that the difference between $R_\nu$ and $R_{\rm gain}$ is small compared with the scale of the system.
We further assume $\dot{Y}_e=0$ at the neutrinosphere.
Thus, for fixed $R_\nu,\rho_0$ and the luminosities and average energies of individual neutrinos $L_{\nu_i}, \epsilon_{\nu_i}$, we obtain the boundary conditions at $r=R_\nu(=R_{\rm gain})$ for temperature and electron fraction at the neutrinosphere $T_0,Y_{e,0}$ such that the conditions $\dot{q}=0$ and $\dot{Y}_e=0$ are fulfilled simultaneously.
The boundary condition for velocity $u_0$ is set to be such that the wind solution becomes transonic.

\subsection{Heating and Cooling Rates and Reaction Rates due to Weak Interactions} 
\label{sec:heating-and-cooling-rates and-reaction-rates-due-to-weak-interactions}
The heating rate due to neutrino and antineutrino absorptions by free nucleons is given by
\begin{eqnarray}
\dot{q}_{\nu N}&\approx&9.84\ N_{\rm A}\ [(1-Y_e)L_{\nu_e,51}\varepsilon_{\nu_e,{\rm MeV}}^2 + Y_e L_{\bar{\nu}_e,51}\varepsilon_{\bar{\nu}_e,{\rm MeV}}^2]\nonumber\\
&\times&\frac{1-g_1(r)}{R_{\nu,6}^2}\Phi(r)^6\ {\rm MeV\ s^{-1}\ g^{-1}},\label{eq:q1}
\end{eqnarray}
where $N_{\rm A}$ is the Avogadro constant, $L_{\nu_i,51}$ is the luminosity of neutrino or antineutrino of flavor $i$ in units of $10^{51}\ {\rm erg\ s^{-1}}$, $\varepsilon_{\nu_i,{\rm MeV}}$ is the average neutrino or antineutrino energy of flavor $i$ in MeV defined by $\varepsilon_{\nu_i} =\sqrt{\langle E_{\nu_i}^3 \rangle/\langle E_{\nu_i} \rangle}$ ($\langle E_{\nu_i}^n \rangle$ denotes the $n$th moment of the energy distribution of the neutrino of flavor $i$), and $R_{\nu,6}$ is the radius of the neutrinosphere in units of $10^6\ {\rm cm}$.
In deriving the above expression, we approximate that all radii of the neutrinospheres of individual neutrino flavors are the same.
The factor $[1-g_1(r)]$ represents the effect that we see the neutrinosphere in a finite solid angle considering the effect of bending neutrino trajectory \citep{2000ApJ...533..424O}, and $g_1(r)$ is defined by
\begin{eqnarray}
g_1(r) = \sqrt{1-\left(\frac{R_\nu}{r}\right)^2\frac{1-2M/r}{1-2M/R_\nu}},
\end{eqnarray}
and $\Phi(r)$ is the redshift factor given by 
\begin{eqnarray}
\Phi(r) = \sqrt{\frac{1-2M/R_\nu}{1-2M/r}}.
\end{eqnarray}

The heating rate due to neutrino-electron scattering is given by
\begin{eqnarray}
\dot{q}_{e\nu}&\approx&2.17\ N_{\rm A}\ \frac{T_{\rm MeV}^4}{\rho_8}\nonumber\\
&&\times \biggl(L_{\nu_e,51}\epsilon_{\nu_e,{\rm MeV}}+L_{\bar{\nu}_e,51}\epsilon_{\bar{\nu}_e,{\rm MeV}}+\frac{6}{7}L_{\nu_x,51}\epsilon_{\nu_x,{\rm MeV}}\biggr)\nonumber\\
&&\times\ \frac{1-g_1(r)}{R_{\nu,6}^2}\Phi(r)^5\ {\rm MeV\ s^{-1}g^{-1}},\label{eq:q3}
\end{eqnarray}
where $L_{\nu_x}$ is the sum of the luminosity of all neutrino flavors except for electron neutrinos and electron antineutrinos, $T_{\rm MeV}$ is the temperature of the fluid in MeV, $\rho_8$ is the mass density in units of $10^8\ {\rm g\ cm^{-3}}$, and $\epsilon_{\nu_i}$ is defined by $\epsilon_{\nu_i} =\langle E_{\nu_i}^2 \rangle/\langle E_{\nu_i} \rangle$ (here $\nu_x$ means the neutrinos except for electron neutrinos and electron antineutrinos). In the following, we choose the relation between $\varepsilon_{\nu_i}$ and $\epsilon_{\nu_i}$ as $\varepsilon_{\nu_i}^2 \approx 1.14 \epsilon_{\nu_i}^2$ as in \cite{1996ApJ...471..331Q}.

The heating rate due to neutrino-antineutrino pair annihilation into electron-positron pairs is given by
\begin{eqnarray}
\dot{q}_{\nu\bar{\nu}}&\approx&12.0\ N_{\rm A}\nonumber \\
&&\times\biggl[ L_{\nu_e,51}L_{\bar{\nu}_e,51}(\epsilon_{\nu_e,{\rm MeV}}+\epsilon_{\bar{\nu}_e,{\rm MeV}})+\frac{6}{7} L_{\nu_x,51}^2\epsilon_{\nu_x,{\rm MeV}}\biggr]\nonumber\\
&&\times\ \frac{g_2(r)}{\rho_8 R_{\nu,6}^4}\Phi(r)^9\ {\rm MeV\ s^{-1}g^{-1}},\label{eq:q5}
\end{eqnarray}
where $g_2$ is defined using $g_1$ as
\begin{equation}
g_2(r) =[1-g_1(r)]^4[g_1(r)^2+4g_1(r)+5].
\end{equation}

The cooling rate due to electron and positron capture by free nucleons $\dot{q}_{eN}$ and the cooling rate due to electron-positron pair annihilation into neutrino-antineutrino pairs $\dot{q}_{e^+e^-}$ are calculated using the expression in \cite{1985ApJ...293....1F} (hereafter FFN85).
The cooling rate per unit time per unit mass due to electron and positron capture by free nucleons is given as
\begin{eqnarray}
q_{eN} &=& m_eN_{\rm A}\frac{\ln 2}{(ft)} \left(\frac{k_{\rm B}T}{m_e}\right)^6 \nonumber \\
&&\times\ \biggl[\left.\frac{(1-Y_e)J}{1-\exp(\eta_\nu^F - \zeta_n - \eta_e^F)}\right|_{\rm pc}\nonumber\\
&&\ +\ \left.\frac{Y_eJ}{1-\exp(\eta_\nu^F - \zeta_n - \eta_e^F)}\right|_{\rm ec}\biggr]\nonumber\\
&\approx&1.96\times10^{-2}\ {\rm MeV\ s^{-1}\ g^{-1}}\  N_{\rm A} T_{\rm MeV}^6\nonumber\\
&&\times\ \biggl[\left.\frac{(1-Y_e)J}{1-\exp(\eta_\nu^F - \zeta_n - \eta_e^F)}\right|_{\rm pc}\nonumber\\
&&\ +\ \left.\frac{Y_eJ}{1-\exp(\eta_\nu^F - \zeta_n - \eta_e^F)}\right|_{\rm ec}\biggr],
\label{eq:eN-cooling-rate}
\end{eqnarray}
where subscripts ``ec" and ``pc" denote electron capture and positron capture, respectively.  Here $m_e$ is the electron mass, $(ft)\approx1.02\times 10^3\ {\rm s}$ is the $ft$-value of the transition between neutrons and protons, $\eta_e^F = \mu_e/k_{\rm B}T$ and $\eta_\nu^F = \mu_\nu/k_{\rm B}T$ are the chemical potentials of electrons (positrons) and electron neutrinos (electron antineutrinos) divided by $k_{\rm B}T$ in the case of electron (positron) capture by free protons (neutrons), and $\zeta_n$ is $-\Delta/k_{\rm B}T$ in the case of electron capture by protons and is $\Delta/k_{\rm B}T$ in the case of positron capture by neutrons ($\Delta=m_n-m_p \approx 1.293\ {\rm MeV}$ is the mass difference between neutron and proton).
$J$ is so called the phase-space factor of this reaction rate and is defined in appendix \ref{app:A}.

The cooling rate due to the pair annihilation of electrons and positrons into a neutrino-antineutrino pair is calculated using the expression in \cite{1986ApJ...309..653C}. According to their expression, the cooling rate is given as
\begin{eqnarray}
\dot{q}_{e^+e^-} &=&\frac{1}{\rho}\frac{G_{\rm F}^2 (D_e + D_\mu + D_\tau)}{9\pi^5} (k_{\rm B}T)^9K\nonumber\\
&\approx&5.43\times10^{-4}\ \frac{T_{\rm MeV}^9}{\rho_8} K\ {\rm MeV\ s^{-1}\ g^{-1}},
\end{eqnarray}
where $G_{\rm F}$ is the Fermi coupling constant, $D_e$, $D_\mu$, and $D_\tau$ are given using the Weinberg angle $\theta_{\rm W}$ as $D_e=1+ 4\sin^2\theta_{\rm W} + 8\sin^4\theta_{\rm W}$, $D_{\mu}=D_{\tau}=1- 4\sin^2\theta_{\rm W} + 8\sin^4\theta_{\rm W}$ and we use the value $\sin^2\theta_{\rm W} = 0.23$ in evaluating these expressions, and $K$ is the phase-space factor of this cooling rate, which is also given in appendix \ref{app:A}.

The reaction rates of electron and positron capture by free nucleons $\lambda_{\rm ec}$ and $\lambda_{\rm pc}$ are given using the expression in FFN85. In their expression, the rates are given by
\begin{eqnarray}
\lambda_{\rm ec}&=&\frac{\ln 2}{(ft)}\left(\frac{k_{\rm B}T}{m_e}\right)^5\left.\frac{I}{1-\exp(\eta_\nu^F - \zeta_n - \eta_e^F)}\right|_{\rm ec}\nonumber\\
&\approx&1.96\times 10^{-2}\ T_{\rm MeV}^5\left.\frac{I}{1-\exp(\eta_\nu^F - \zeta_n - \eta_e^F)}\right|_{\rm ec}{\rm s^{-1}},\hspace{7mm}\\
\lambda_{\rm pc}&=&\frac{\ln 2}{(ft)}\left(\frac{k_{\rm B}T}{m_e}\right)^5\left.\frac{I}{1-\exp(\eta_\nu^F - \zeta_n - \eta_e^F)}\right|_{\rm pc} \nonumber\\
&\approx&1.96\times 10^{-2}\ T_{\rm MeV}^5\left.\frac{I}{1-\exp(\eta_\nu^F - \zeta_n - \eta_e^F)}\right|_{\rm pc}{\rm s^{-1}},
\end{eqnarray}
where $(ft)$, $\eta_\nu^F$, $\eta_e^F$, and $\zeta_n$ are defined in (\ref{eq:eN-cooling-rate}), and $I$ is the phase-space factor of this reaction rate, which is defined in appendix \ref{app:A}.
In the following, we set $\eta_\nu^F=0$.

The rates of neutrino and antineutrino absorption by free nucleons $\lambda_{\nu n}$ and $\lambda_{\bar{\nu}p}$ are given using the expression in \cite{2000ApJ...533..424O} as
\begin{eqnarray}
\lambda_{\bar{\nu}p}&\approx& 9.84  \, \left( \epsilon_{\bar{\nu}_e,{\rm MeV}} - 2\Delta_{\rm MeV} + 1.2\frac{\Delta_{\rm MeV}^2}{\epsilon_{\bar{\nu}_e,{\rm MeV}}} \right)\nonumber\\
&&\times\ \frac{L_{\bar{\nu}_e,51}\Phi(r)^6[1-g_1(r)]}{R_{\nu,6}^2} \,{\rm s^{-1}}, \label{eq:nub-p}\\
\lambda_{\nu n}&\approx& 9.84  \, \left( \epsilon_{\nu_e,{\rm MeV}} + 2\Delta_{\rm MeV} + 1.2\frac{\Delta_{\rm MeV}^2}{\epsilon_{\nu_e,{\rm MeV}}} \right)\nonumber \\
&&\times\ \frac{L_{\nu_e,51}\Phi(r)^6[1-g_1(r)]}{R_{\nu,6}^2} \,{\rm s^{-1}}\label{eq:nu-n}.
\end{eqnarray}

\subsection{The Parameters of the Wind}
The parameters of the winds are the luminosities and average energies of the neutrinos of each flavor $L_{\nu_e},L_{\bar{\nu}_e},L_{\nu_x},\epsilon_{\nu_e},\epsilon_{\bar{\nu}_e},\epsilon_{\nu_x}$, the mass of the neutron star $M$, the radius of the neutrinosphere $R_\nu$, which is assumed to be the gain radius $R_{\rm gain}$, and the mass density at the gain radius $\rho_0$.

Here we explain the fiducial parameter set and parameter range adopted in this study.
We set the parameters on the basis of the numerical relativity simulation of the collapse of a $100M_\odot$ star in \cite{2012PTEP.2012aA304S}.
They showed that in the center of the collapsing star an mPNS with the gravitational mass of $M\approx 3M_\odot$ and the neutrinosphere with the radius of $R_\nu\approx 15\ {\rm km}$ is temporarily formed.
Within its lifetime, on order of seconds, the neutrino luminosity and the mass of the neutron star become nearly constant in time.
In this phase, the neutrino luminosity of the neutron star is $L_{\nu} \sim 10^{53}\ {\rm erg\ s^{-1}}$, the ratio of the luminosity of the electron antineutrino to that of the electron neutrino is $L_{\bar{\nu}_e}/L_{\nu_e} \approx 1.1$, and the ratio of the luminosities of $\mu$ and $\tau$ neutrinos to the electron neutrino luminosity is  $L_{\nu_x}/L_{\nu_e} \approx 0.1$.
Thus, we consider the steady flow in this phase under these conditions $M=3 M_\odot$, $R_\nu = 15\ {\rm km}$, $L_{\nu_e} =1.5\times 10^{53}\ {\rm erg\ s^{-1}}$, $L_{\bar{\nu}_e}/L_{\nu_e} = 1.1$, $L_{\nu_x}/L_{\nu_e} = 0.1$ from the numerical relativity simulation.

The average energies of the electron neutrinos $\epsilon_{\nu_e}$ and electron antineutrinos $\epsilon_{\bar{\nu}_e}$ are very important because they determine the electron fraction $Y_e$ and, consequently, the yield of nucleosynthesis.
These values have some uncertainty caused by the approximate treatment of the neutrino transport and the finite density effects on neutrino opacities \citep[e.g.,][]{2012PhRvC..86f5806H,2012PhRvC..86f5803R}.
In the numerical simulation \citep{2012PTEP.2012aA304S}, their average energies are $\epsilon_{\nu_e}\approx 11\ {\rm MeV}$ and $\epsilon_{\bar{\nu}_e}\approx 16\ {\rm MeV}$.
Thus, we construct the wind solutions and calculate nucleosynthesis on trajectories of the wind for the energy region $\epsilon_{\nu_e}=9-16\ {\rm MeV}$ and $\epsilon_{\bar{\nu}_e}=11-18\ {\rm MeV}$, respectively.
We also set the fiducial parameters of $\epsilon_{\nu_e}$ and $\epsilon_{\bar{\nu}_e}$ as 11 and 16 MeV as in \cite{2012PTEP.2012aA304S}.
The average energies of the other flavors of neutrinos, $\mu$ and $\tau$ neutrino and its antiparticles, $\epsilon_{\nu_x}$, is less important in constructing the wind solutions or nucleosynthesis results because the luminosity of them is much smaller than that of electron and electron antineutrinos.
Thus, they are set to be, from the simulation, $25\ {\rm MeV}$ for all wind solutions.

We set the mass density at the neutrinosphere $\rho_0$ to be $10^{11}\ {\rm g\ cm^{-3}}$.
This value is 10 times as large as the mass density in the neutrino-driven wind models of an ordinary SN in \cite{2000ApJ...533..424O} and \cite{2001ApJ...554..578W}, where it is set as $\rho_0=10^{10}\ {\rm g\ cm^{-3}}$.
This is due to the very large neutrino luminosity in our model, which is about 100 times larger than their models.
At the surface of the PNS, dominant heating and cooling processes are neutrino absorption reactions by free nucleons and their inverse reactions (\ref{reac:nup}) and (\ref{reac:nun}), and the balance between heating and cooling gives the relation among the neutrino luminosity $L_\nu$, the average energy of the neutrinos $\epsilon_\nu$, and the average energy of the electrons and positrons captured by nucleons $\bar{E}_e$ as
\begin{eqnarray}
L_\nu \epsilon_\nu^2 \propto \bar{E}_e^6 \label{eq:lumi-den}.
\end{eqnarray}
The average electron energy at the surface of the PNS $\bar{E}_e$ is almost equal to the chemical potential of electrons $\mu_e$ because of the electron degeneracy.
The average energy of neutrinos can be estimated as the surface temperature $T$.
Using the relation $\mu_e \propto \rho^{1/3}$ for relativistic degenerate electrons, Equation (\ref{eq:lumi-den}) can be rewritten as
\begin{eqnarray}
L_\nu \propto \mu_e^6 T^{-2} \propto \rho^2 T^{-2}.
\end{eqnarray}
Assuming that the surface temperatures are almost the same between the ordinary PNS and the mPNS, the surface density of the mPNS is about 10 times larger than that of the ordinary PNS.

Summarizing the above, we give the parameter range of the wind in this work, from the simulation, in Table \ref{tab:modelpar}.

\subsection{Comparison of Wind Solutions}
Here we discuss the differences between our mPNS wind and the ordinary PNS wind.
The mass of the mPNS in our model is $\approx 3M_\odot$, and its radius (the radius of the neutrinosphere) is about $\approx15\ {\rm km}$.
On the other hand, the mass of the ordinary PNS is $\approx 1.4M_\odot$ and its radius is about $\approx 10\ {\rm km}$.
Thus, the mPNS in our model has a compactness parameter ($C\equiv M/R \approx 0.3$) larger than that of the ordinary PNS ($C\approx 0.2$).
It indicates that general relativistic effects become more important in the mPNS wind than that in the ordinary PNS wind.
In addition, the neutrino luminosity in the mPNS model is about $\sim 10^{53}\ {\rm erg\ s^{-1}}$, which is about 100 times larger than that in the ordinary PNS (about $10^{51}\ {\rm erg\ s^{-1}}$).

We solve the wind equations, i.e., Equations (\ref{eq:u-evol})-(\ref{eq:Ye-evol}), with the two parameter sets given in Table \ref{tab:modelpar}.
One is our fiducial parameter set of the mPNS model, and the other is the parameter set of the ordinary PNS \citep[the average energies of neutrinos are determined based on][]{2013ApJ...770L..22W}.
We discuss the differences in the entropy, the velocity, and the temperature between them.

\begin{table*}[htbp]
\caption{The Range of Parameters, the Fiducial Parameter Set of the mPNS Wind, and Those of the Ordinary PNS Wind}
\centering
\begin{tabular*}{\hsize}{@{\extracolsep{\fill}}lccccccccc}
\hline \hline
Model & $M$&$R_{\nu}$&$\rho_{0}$&$L_{\nu_e} $&$L_{\bar{\nu}_e}$&$L_{\nu_x}$&$\epsilon_{\nu_e}$&$\epsilon_{\bar{\nu}_e}$&$\epsilon_{\nu_x}$ \\
&$(M_\odot)$ & $({\rm km})$ & $({\rm g\ cm^{-3}})$ & \multicolumn{3}{c}{$({\rm 10^{51}erg\ s^{-1}})$} & \multicolumn{3}{c}{$({\rm MeV})$} \\ \hline
Parameter range& 3 & 15 &$10^{11}$& 150 & 165 & 15 & 9-16 & 11-18 & 25\\
Fiducial & 3 & 15 &$10^{11}$ & 150 & 165 & 15 & 11 & 16 & 25\\
Ordinary PNS & 1.4 & 10 & $10^{10}$ & 1 & 1 & 1 & 12 & 14 & 14\\
\hline
\end{tabular*}
\label{tab:modelpar}
\end{table*}

\begin{figure}[htbp]
 \begin{center}
  \includegraphics[width=0.9\hsize]{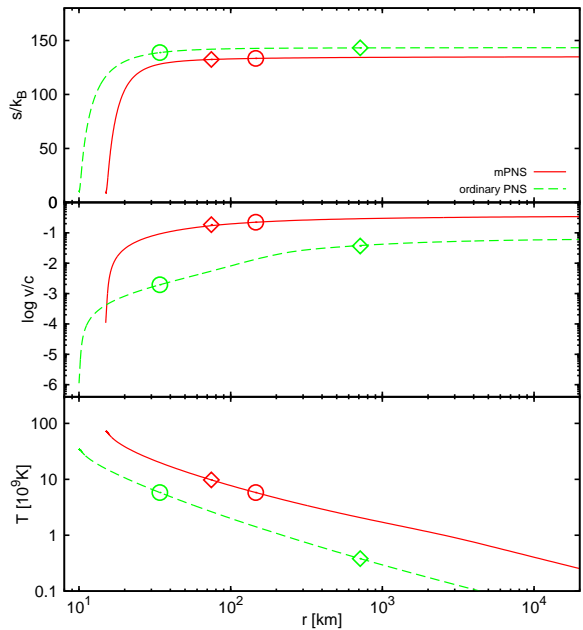}
 \end{center}
\caption{Radial profiles of entropy (top), velocity (middle), and temperature (bottom) for the two wind trajectories. The wind of the mPNS model is shown by solid lines and the wind of the ordinary PNS is shown by the dashed lines. The important radii are marked with symbols, including the critical point ($diamond$) and the radius of alpha-particle formation (the point of $T=0.5\ {\rm MeV}=5.8\times10^9\ {\rm K}$, $circle$).}
\label{fig:compare}
\end{figure}

Figure \ref{fig:compare} shows the radial profiles of the entropy, the velocity, and the temperature for the wind trajectories of the mPNS wind and the ordinary PNS wind.
The terminal entropy of the mPNS wind, which is about 130-140 (here the term ``entropy" means the entropy per baryon in units of $k_{\rm B}$), is almost the same as that of the ordinary PNS wind (see top panel of Figure \ref{fig:compare}).
Qualitatively, this can be explained as follows.
The change of the entropy per baryon, $\Delta s$, of the wind may be given as
\begin{equation}
\Delta s \propto \int \frac{dq}{T} = \frac{\Delta Q}{\bar{T}},\label{eq:delta-s}
\end{equation}
where $\Delta Q = \int dq$ is the total heat obtained by the fluid, and $\bar{T}^{-1} = (\Delta Q)^{-1}\int dq\ T^{-1}$ is the mean temperature weighted by $\dot{q}$.
$\Delta Q$ will be of the order of the gravitational potential at the PNS surface $\sim M/R_\nu=C$.
Therefore, the wind blown off from the mPNS, which has larger compactness parameter $C$, is expected to have larger entropy than that of the wind blown off from the ordinary PNS.
However, the change of the entropy of the mPNS wind is suppressed by higher $\bar{T}$ than that of the ordinary PNS wind (see bottom panel of Figure \ref{fig:compare}) because of higher neutrino luminosity.
These two opposite effects cancel each other, and as a result, the terminal entropies of the mPNS and the ordinary PNS winds have similar values.

The biggest difference between the properties of the mPNS wind and the ordinary PNS wind is the temperature-decrease timescale at $T=0.5\ {\rm MeV}$, denoted by $\tau_{T,0.5}$.
It is defined as
\begin{equation}
\tau_{T,0.5} = \int_{T=0.5\ {\rm MeV}}^{T=0.5/e\ {\rm MeV}} d\tau \label{eq:t-tau},
\end{equation}
where $e$ is the base of natural logarithm \citep[this definition is based on][]{2000ApJ...533..424O}.
When the temperature decreases below $T\approx 0.5\ {\rm MeV}$, free protons and neutrons are reassembled into alpha particles so $\tau_{T,0.5}$ represents the timescale in which the seed elements (the elements heavier than ${}^{12}{\rm C}$) are produced from alpha particles.
Therefore, the temperature-decrease timescale at the point is the useful quantity that represents the inefficiency of the production of seed elements through triple-$\alpha$ and ${}^{4}{\rm He}(\alpha n,\gamma){}^{9}{\rm Be}(\alpha,n){}^{12}{\rm C}$ reactions \citep[e.g.,][]{1997ApJ...482..951H}.

The bottom panel of Figure \ref{fig:compare} shows the temperature of the mPNS and the ordinary SN as a function of the radius.
We see that, in the ordinary PNS wind, the radius of the critical point, indicated by a diamond, is larger than the radius where the temperature drops to 0.5 MeV, i.e., the alpha-particle formation radius.
On the other hand, in the mPNS wind, the relation of these radii is opposite and the temperature decreases to 0.5 MeV after the velocity reaches its asymptotic value (see also the middle panel).
Therefore, the temperature-decrease timescale at $T=0.5\ {\rm MeV}$ of the mPNS wind is expected to be shorter than that of the ordinary PNS wind.
The value of $\tau_{T,0.5}$ of the mPNS wind is about 6 ms and indeed much shorter than that of the ordinary PNS ($\approx 40-50\ {\rm ms}$).
This trend can be understood more quantitatively as follows.
We define the temperature-decrease timescale at any temperature $\tau_T$ as
\begin{eqnarray}
\tau_T^{-1} &=& -\ \frac{1}{T}\frac{dT}{d\tau}\nonumber\\
&\approx& -\ \frac{1}{3\rho}\frac{d\rho}{d\tau}\nonumber\\
&=& \frac{1}{3}\left(\frac{2u}{r} + \frac{1}{u}\frac{du}{d\tau}\right). \label{eq:t-tau-approx}
\end{eqnarray}
In this equation, we assume $dT/d\tau <0$ and,  from the first to the second line, we use the fact that the radiation component dominates most of the total entropy and that the entropy is almost constant in time in all regions where $T\lesssim 10^{10}\ {\rm K}$ (see top panel of Figure \ref{fig:compare}).
From the second to the third equation, we use the proper-time derivative of Equation (\ref{eq:wind-mdot}).
The value $\tau_{T,0.5}$ represents the time during which the temperature decreases from $0.5\ {\rm MeV}$ to $0.5/e\ {\rm MeV}$, and thus it can be written as $\tau_T(T=0.5\ {\rm MeV})$ approximately.
As seen in the middle panel of Figure \ref{fig:compare}, the velocity of the fluid at $T=0.5\ {\rm MeV}$ (open circle shown in the figure) is about $u\approx 0.22c$ and the radius is about $r\approx 140\ {\rm km}$ in the mPNS wind, while these values of the ordinary PNS wind are $u \approx 2.4\times 10^{-3}c$ and $r \approx 35\ {\rm km}$.
Therefore, the first term in the parentheses of Equation (\ref{eq:t-tau-approx}) of the mPNS wind, which is dominant in the parentheses, is about one order larger than that of the ordinary PNS.

Fewer seed elements are produced as $\tau_{T,0.5}$ becomes shorter, and then the production of larger mass number elements in the subsequent process after the seed element production becomes more efficient.
Thus, while the entropy is almost the same, the circumstance of the mPNS wind of short temperature-decrease timescale is more attractive for heavy-element synthesis than that of the ordinary PNS wind.

\section{Nucleosynthesis calculation}
\label{sec:nucleosynthesis-calculation}
The nucleosynthesis calculation for the neutrino-driven winds in the mPNS is started when the temperature of the fluid element decreases to $T_9=9$, where $T_9$ is the temperature in units of $10^9\ {\rm K}$.
This temperature is still high enough for the system to be the nuclear statistical equilibrium (NSE).
At that point, the fluid material is mainly composed of free nucleons under NSE.
Therefore, the initial composition of the fluid material is taken to be $X_p = Y_{e,9}$ and $X_n = 1-Y_{e,9}$, where $Y_{e,9}$ is the initial electron fraction for the nucleosynthesis calculation, which is set as the electron fraction when the temperature of the wind becomes $T_9=9$.

We use a nuclear reaction network for our nucleosynthesis calculations.
The network consists of 5406 isotopes (see Table \ref{tab:nuclist}) extended from \cite{2011MNRAS.412L..78Y} and \cite{2010AIPC.1269..339S}.
We include all relevant reactions, i.e., thermonuclear reactions and their inverse reactions, and weak interactions such as $\beta$-decays and electron or positron captures.
Thermonuclear reaction rates are taken from the JINA REACLIB database \citep{2010ApJS..189..240C}.
We include the expression of \cite{1973ApJ...181..457G} and \cite{1979ApJ...234.1079I} for the correction for thermonuclear reaction rates due to the electron screening.
Weak reaction rates are taken from \cite{2001ADNDT..79....1L}, \cite{1994ADNDT..56..231O}, \cite{1987ADNDT..36..375T}, and FFN85 for the reactions whose dependence on the temperature and the density is given.
We follow \cite{Horiguchi+(1996)} for weak reaction rates whose dependence on them is not given, although the experimental rates are known.
For the nuclei that lack experimental decay-rate information, we use Tachibana (2000, private communication) for $\beta^+$-decay rates and \cite{2003PhRvC..67e5802M} for $\beta ^-$-decay rates.
In addition, we include neutrino absorption reactions of free nucleons using Equations (\ref{eq:nub-p}) and (\ref{eq:nu-n}).

\begin{table*}[t]
\label{tab:nuclist}
\caption{
A List of Nuclei in Our Nuclear Reaction Network.
}
\centering
\scriptsize
\begin{tabular*}{\hsize}{@{\extracolsep{\fill}}lllllllllllllll}
\hline\hline
Element&$A_{\rm min}$&$A_{\rm max}$ &Element&$A_{\rm min}$&$A_{\rm max}$ &Element&$A_{\rm min}$&$A_{\rm max}$ &Element&$A_{\rm min}$&$A_{\rm max}$ &Element&$A_{\rm min}$&$A_{\rm max}$\\ \hline
 n &    1 &    1 &   Ar   &   27 &   67 &   Kr   &   63 &  124 &  Xe    &  103 &  182 &  Hf    &  149 &  240\\
 H &    1 &    3 &    K   &   29 &   70 &   Rb   &   66 &  128 &  Cs    &  106 &  185 &  Ta    &  151 &  243\\
 He  &    3 &    6 &   Ca   &   30 &   73 &   Sr   &   68 &  131 &  Ba    &  108 &  189 &   W    &  154 &  247\\
   Li  &    6 &    9 &   Sc   &   32 &   76 &    Y   &   70 &  134 &  La    &  110 &  192 &  Re    &  156 &  250\\
   Be  &    7 &   12 &   Ti   &   34 &   80 &   Zr   &   72 &  137 &  Ce    &  113 &  195 &  Os    &  159 &  253\\
    B  &    8 &   14 &    V   &   36 &   83 &   Nb   &   74 &  140 &  Pr    &  115 &  198 &  Ir    &  162 &  256\\
    C  &    9 &   18 &   Cr   &   38 &   86 &   Mo   &   77 &  144 &  Nd    &  118 &  201 &  Pt    &  165 &  260\\
   N   &   11 &   21 &   Mn   &   40 &   89 &   Tc   &   79 &  147 &  Pm    &  120 &  205 &  Au    &  167 &  263\\
   O   &   13 &   22 &   Fe   &   42 &   92 &   Ru   &   81 &  150 &  Sm    &  123 &  208 &  Hg    &  170 &  266\\
   F   &   14 &   26 &   Co   &   44 &   96 &   Rh   &   83 &  153 &  Eu    &  125 &  211 &  Tl    &  173 &  269\\
  Ne   &   15 &   41 &   Ni   &   46 &   99 &   Pd   &   86 &  156 &  Gd    &  128 &  214 &  Pb    &  175 &  273\\
  Na   &   17 &   44 &   Cu   &   48 &  102 &   Ag   &   88 &  160 &  Tb    &  130 &  218 &  Bi    &  178 &  276\\
  Mg   &   19 &   47 &   Zn   &   51 &  105 &   Cd   &   90 &  163 &  Dy    &  133 &  221 &  Po    &  182 &  276\\
  Al   &   21 &   51 &   Ga   &   53 &  108 &   In   &   92 &  166 &  Ho    &  136 &  224 &  At    &  186 &  279\\
  Si   &   22 &   54 &   Ge   &   55 &  112 &   Sn   &   94 &  169 &  Er    &  138 &  227 &        &     &    \\
   P   &   23 &   57 &   As   &   57 &  115 &   Sb   &   97 &  172 &  Tm    &  141 &  230 &       &     &   \\
   S   &   24 &   60 &   Se   &   59 &  118 &   Te   &   99 &  176 &  Yb    &  143 &  234 &       &     & \\
  Cl   &   26 &   63 &   Br   &   61 &  121 &   I    &  101 &  179 &  Lu    &  146 &  237 &      &    & \\ 
  \hline
\end{tabular*}
\begin{tablenotes}
\footnotesize
\item {\bf Note.}\\$^{5}$He, $^{8}$Be, and $^{9}$B are not included in the network.
\end{tablenotes}
\end{table*}

\section{Results}
\label{sec:results}
\subsection{The Properties of the Wind Trajectories}
\begin{figure}[htbp]
\begin{center}
\includegraphics[width=\hsize]{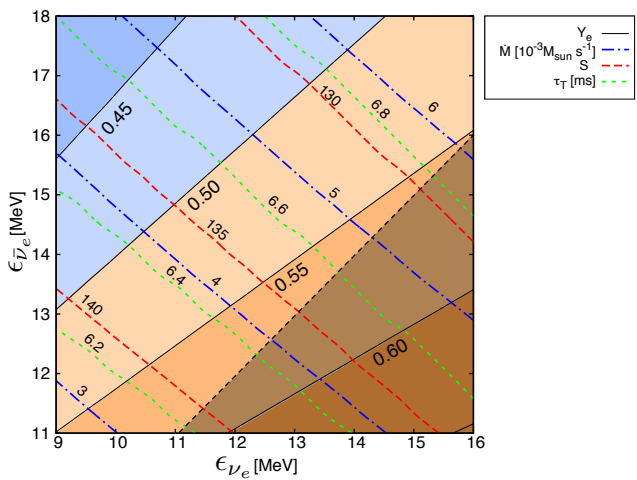}
\end{center}
\caption{
Contours of the electron fraction, the temperature-decrease timescale, the entropy, and the mass outflow rate as functions of $\epsilon_{\nu_e}$ and $\epsilon_{\bar{\nu}_e}$.
The other parameters are adopted from Table \ref{tab:modelpar}.
The black solid lines indicate the electron fraction at the beginning of the nucleosynthesis calculation $Y_{e,9}$.
The green dotted lines indicate the temperature-decrease timescale, $\tau_T$ at $T=0.5\ {\rm MeV}$ in units of ms.
The red dashed lines indicate the entropy per baryon $s$ at $T=0.5\ {\rm MeV}$ in units of $k_{\rm B}$.
The blue dot-dashed lines indicate the mass outflow rate $\dot{M}$ in units of $10^{-3}M_\odot\ {\rm s^{-1}}$.
The region colored in blue is the region where $Y_{e,9}<0.5$, and the region in orange is the region where $Y_{e,9}>0.5$.
The regions where the electron fraction deviates from 0.5 have deeper color.
The shaded region is the region where $\epsilon_{\nu_e}>\epsilon_{\bar{\nu}_e}$ and thus is an unrealistic parameter range.
}
\label{fig:fiducial_wind}
\end{figure}

First, we discuss the properties of constructed winds.
In Figure \ref{fig:fiducial_wind}, we show the contours of the electron fraction at the beginning of the nucleosynthesis calculation ($T_9=9$) $Y_{e,9}$, the entropy $s$ at $T=0.5\ {\rm MeV}$, and the temperature-decrease timescale at $T=0.5\ {\rm MeV}$, $\tau_{T,0.5}$, which are important for the nucleosynthesis.

The electron fraction at the beginning of the nucleosynthesis calculation $Y_{e,9}$, which is set from the wind solutions, has a wide variation from $\approx0.40-0.65$ in our parameter region.
It becomes 0.50 when $\epsilon_{\nu_e}$ is about 4 MeV smaller than $\epsilon_{\bar{\nu}_e}$.
A larger difference between $\epsilon_{\nu_e}$ and $\epsilon_{\bar{\nu}_e}$ provides $Y_{e,9}<0.5$, i.e., neutron-rich wind, and a smaller difference provides proton-rich wind.
As shown in the next subsection, the electron fraction strongly affects the abundance distribution of the wind material.

The obtained electron fraction is close to the value $Y_{e,{\rm eq}}$ from the balance of the two reactions: $p+\bar{\nu}_e\rightarrow n + e^+$ and $n+\nu_e\rightarrow p + e^-$.
The value $Y_{e,{\rm eq}}$ is given approximately from Equation (77) in \cite{1996ApJ...471..331Q}.

The entropy of the wind is distributed in the range $s\approx 125-145$.
It gets smaller as the average energy of the neutrinos gets larger.
This can be understood from the equation of the entropy change, i.e., Equation (\ref{eq:delta-s}).
As the average energy of neutrinos becomes larger, the heating rate and $\bar{T}$ become larger, so the entropy increase due to the heating by neutrino absorption is suppressed by the high temperature.
On the other hand, $\Delta Q$ is determined by the gravitational binding energy of PNSs and is almost independent of the average energy of neutrinos.
Thus, the entropy is smaller when the average energy of neutrinos is larger.

The temperature-decrease timescale at $T=0.5\ {\rm MeV}$, $\tau_{T,0.5}$, which is defined as Equation (\ref{eq:t-tau}), slightly decreases (from 6 to 7 ms) as the average energies of the neutrinos decrease.
This can be understood using the last line of Equation (\ref{eq:t-tau-approx}).
It is sufficient to pay attention to the first term, which is proportional to $u/r|_{T=0.5\ {\rm MeV}}$.
This is because the radius where the temperature becomes $0.5\ {\rm MeV}$ is larger than the radius of the critical point and thus the second term, $u^{-1} du/d\tau = du/dr$, is subdominant compared to the first term.
The radius where the temperature becomes $0.5\ {\rm MeV}$ gets larger as $\epsilon_{\nu_e}$ or $\epsilon_{\bar{\nu}_e}$ becomes large because of more efficient heating.
On the other hand, the terminal velocity is independent of $\epsilon_{\nu_e}$ or $\epsilon_{\bar{\nu}_e}$  because it is determined by the gravitational binding energy at the launching point of the fluid.
Therefore, the winds having small $\epsilon_{\nu_e}$ or $\epsilon_{\bar{\nu}_e}$ have small $\tau_{T,0.5}$ because they have large $r|_{T=0.5\ {\rm MeV}}$.

The mass outflow rate $\dot{M}$ is on the order of $10^{-3}M_\odot\ {\rm s^{-1}}$ throughout the figure and becomes large as the average energy of neutrinos is large because of the efficient heating.

The contours of the temperature-decrease timescale, the mass outflow rate, and the entropy are almost parallel to each other.
This is because all these quantities depend mainly on the heating rates, i.e., on $\epsilon_{\nu_e}$ and $\epsilon_{\bar{\nu}_e}$ (see Equations (\ref{eq:q1}), (\ref{eq:q3}) and (\ref{eq:q5})).
On the other hand, the electron fraction $Y_{e,9}$ depends on the difference of  the average energies between electron neutrino and electron antineutrino $\epsilon_{\nu_e}-\epsilon_{\bar{\nu}_e}$ \citep[from the approximate form of Equation (77) in][]{1996ApJ...471..331Q}.

The average energies of electron neutrinos and electron antineutrinos usually have the relation $\epsilon_{\nu_e}<\epsilon_{\bar{\nu}_e}$, and the shaded region in Figure \ref{fig:fiducial_wind} is physically less relevant.
This is because PNSs have much neutrons than protons, so the radius of the neutrinosphere of electron neutrinos is larger than that of electron antineutrinos.

\begin{figure}[htbp]
\begin{center}
\includegraphics[width=\hsize]{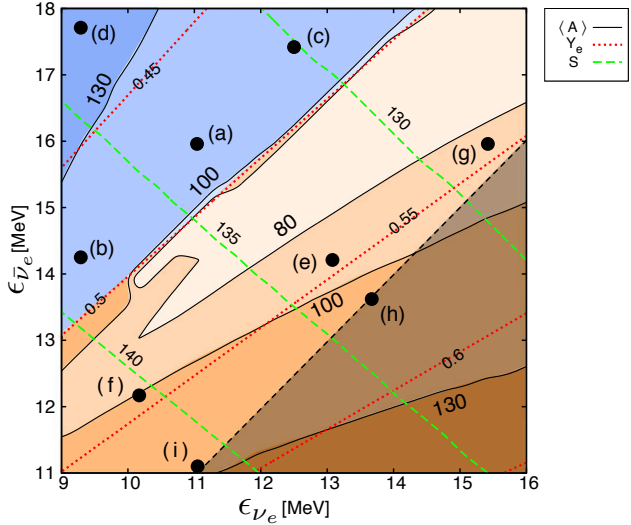}
\end{center}
\caption{
The contours of the average mass number of final abundance distribution $\left<A\right>$, the electron fraction at the beginning of the nucleosynthesis calculation $Y_{e,9}$, and the entropy $s$.
The solid lines indicate the average mass number $\left< A \right>$ of the heavy-element set.
The red doted lines indicate the electron fraction $Y_{e,9}$.
The green dashed lines indicate the entropy $s$ per baryon at $T=0.5\ {\rm MeV}$ in units of $k_{\rm B}$.
The $r$-process and the $\nu p$-process proceed in blue and orange region, respectively.
The deeper the color of a region is, the larger mass number nuclei are produced in the trajectory in the region. The shaded region at the lower right of the black-dashed line covers unrealistic parameter region where $\epsilon_{\nu_e}>\epsilon_{\bar{\nu}_e}$.
}
\label{fig:fiducial}
\end{figure}

\subsection{Dependence of the Abundance Distribution on the Average Energies of Electron Neutrinos and Electron Antineutrinos}
We calculate the nucleosynthesis for the trajectories with parameters listed in Table \ref{tab:modelpar}.
In Figure \ref{fig:fiducial}, three contours for the electron fraction at the beginning of the nucleosynthesis calculation $Y_{e,9}$, the entropy $s$ at $T=0.5\ {\rm MeV}$, and the average mass number of the synthesized heavy elements $\left<A\right>$, defined below, are shown in the $\epsilon_{\nu_e}-\epsilon_{\bar{\nu}_e}$ plane.
In order to see the typical mass number of heavy elements synthesized in each trajectory, we define the {\it heavy-element set} ${\mathcal A_{>48}}$ as the nuclei that have the mass number $A>48$, which includes iron-group and heavier elements, and the average mass number of the heavy-element set of the final abundance distribution $\left<A\right>$ as
\begin{equation}
\left< A \right> = \frac{1}{Y_h}\sum_{i\in{\mathcal A_{>48}}} A_i Y(Z_i,A_i),\label{eq:amean}
\end{equation}
where $Y(Z_i,A_i)$ is the abundance in number of the species having the atomic number $Z_i$ and the mass number $A_i$.
In Equation (\ref{eq:amean}), $Y_h$ is the abundance of the heavy-element set defined as
\begin{equation}
Y_h=\sum_{i\in{\mathcal A_{>48}}}Y(Z_i,A_i).
\end{equation}

Figure \ref{fig:fiducial} is divided into two regions parted on the contour of $Y_{e,9}=0.5$.
In the region where $Y_{e,9}<0.5$ (the blue region in Figure \ref{fig:fiducial}), heavy elements are synthesized through the $r$-process, while in the region where $Y_{e,9}>0.5$ (the orange region in Figure \ref{fig:fiducial}), the $\nu p$-process is the main process that produces heavy elements.

The average mass number $\left< A \right>$ strongly depends on the electron fraction.
In the region where $|Y_{e,9}-0.5|$ is larger, heavier elements are synthesized.
In the region with $Y_{e,9}\lesssim 0.45$, the average mass number exceeds 130 through the $r$-process.
On the other hand, strong $\nu p$-process proceeds in the region with $Y_{e,9}\gtrsim 0.55$ and the average mass number becomes $\left<A\right>\gtrsim 80-100$.
For the synthesis of heavy elements, higher entropy is favored because high entropy prevents seed elements from being produced by strong photo-disintegration and the number of neutrons and protons captured by each seed nucleus in the subsequent phase becomes large.
We see that even for the same electron fraction, the trajectory that has higher entropy has larger average mass number for the above reason.
However, the dependence of the average mass number on the entropy is weaker than the dependence on the electron fraction.

\begin{table*}[t]
\caption{
Properties of Trajectories
}
\centering
\begin{tabular*}{\hsize}{@{\extracolsep{\fill}}crrccccccrl}
\hline \hline
 & $\epsilon_{\nu_e}$ & $\epsilon_{\bar{\nu}_e}$ & $Y_{e,9}$ & $\tau_{T,0.5}$ & $s$  & $\dot{M}$ & $f_{200}$ & $\Delta_n$ & $\left<A\right>$ &Mainly Produced Nuclides\\
 &  \multicolumn{2}{c}{(MeV)} & & (ms) & \multicolumn{3}{c}{$(10^{-3}M_\odot{\rm s^{-1}})$} &&& \\ 
  \hline
(a) & 11.0 & 16.0 & 0.481 & 6.59 & 133  & 4.62 & 0.696 &$\cdots$& 118 & ${}^{122,124}{\rm Sn}$, ${}^{121,123}{\rm Sb}$, ${}^{{\bf125},126,128}{\bf Te}$, ${}^{127}{\rm I}$\\
(b)&   9.29 & 14.2 & 0.482 & 6.35 & 138 & 3.66 & 0.730 &$\cdots$& 123 & ${}^{125,126,128,130}{\rm Te}$, ${}^{127}{\rm I}$, ${}^{{\bf 129},131,132,134}{\bf Xe}$,  ${}^{133}{\rm Cs}$\\
(c) & 12.5 & 17.4 & 0.480 & 6.76 & 130 & 5.54 & 0.676 &$\cdots$& 115 & ${}^{109}{\rm Ag}$,${}^{111,116}{\rm Cd}$, ${}^{122,124}{\rm Sn}$, ${}^{121,123}{\rm Sb}$, ${}^{{\bf 125},126}{\bf Te}$, ${}^{127}{\rm I}$ \\
(d) &   9.29 & 17.7 & 0.421 & 6.60 & 133 & 4.67 & 0.795 &$\cdots$& 139 & ${}^{134}{\rm Xe}$, ${}^{\bf 133}{\bf Cs}$, ${}^{135,137}{\rm Ba}$, ${}^{177,178,179}{\rm Hf}$, ${}^{198}{\rm Pt}$, ${}^{198}{\rm Au}$\\
(e) & 13.1 & 14.2 & 0.545 & 6.59 & 133 & 4.63 & $\cdots$ & 39.7&  92.4 & ${}^{96,{\bf 98},100}{\bf Ru}$, ${}^{102,104}{\rm Pd}$, ${}^{108}{\rm Cd}$\\
(f)  & 10.2 & 12.2 & 0.543 & 6.23 & 141 & 3.33 & $\cdots$ & 49.2& 101 & ${}^{{\bf 98},100}{\bf Ru}$, ${}^{103}{\rm Rh}$, ${}^{102,110}{\rm Pd}$, ${}^{109}{\rm Ag}$, ${}^{108,111,113}{\rm Cd}$, ${}^{115}{\rm In}$\\
(g) & 15.4 & 16.0 & 0.545 & 6.84 & 129 & 5.95 & $\cdots$ & 31.1&  84.4 & ${}^{84}{\rm Sr}$, ${}^{96,{\bf 98}}{\bf Ru}$, ${}^{102,104}{\rm Pd}$, ${}^{108}{\rm Cd}$\\
(h) & 13.7 & 13.6 & 0.566 & 6.59 & 133 & 4.60 & $\cdots$ & 55.5& 106 & ${}^{{\bf 98},100}{\bf Ru}$, ${}^{110}{\rm Pd}$, ${}^{108,111,113,116}{\rm Cd}$, ${}^{115}{\rm In}$, ${}^{114,115}{\rm Sn}$\\
(i)  & 11.0 & 11.0 & 0.587 & 6.18 & 141 & 3.21 & $\cdots$ & 110& 129 & ${}^{\bf 138}{\bf La}$\\ \hline
\end{tabular*}
\begin{tablenotes}
\footnotesize
\item {\bf Note.} In the column ``Mainly Produced Nuclides," we list top 10 nuclides which have the production factors larger than 1/10 times the largest production factor in each trajectory.
The nuclide which has the largest production factor in each trajectory is written in bold.
The parameter $f_{200}$ is defined as Equation (\ref{eq:f200}). 
\end{tablenotes}
\label{tab:trajs}
\end{table*}

\subsection{Details of the Nucleosynthesis Processes}

In the previous subsection, we showed that heavy elements with $A\gtrsim100$ can be synthesized through the $r$-process or the $\nu p$-process in the neutrino-driven winds of the mPNS.
Here we show more details for the average mass number of the heavy-element set and the main products in nine wind trajectories (a)-(f) in Figure \ref{fig:fiducial}.
In Table \ref{tab:trajs}, we list the properties of these wind trajectories and the nucleosynthesis results of them.
Trajectory (a) is the wind with the fiducial parameters (see Table \ref{tab:modelpar}).
In order to see the effect of the entropy, we consider trajectories (b) and (c), which have almost the same $Y_{e,9}$ as trajectory (a) but different $s$ ($130-138$).
Trajectory (d) has almost the same $s$ as trajectory (a) but lower $Y_{e,9}$ of $0.42$.

In order to see the nucleosynthesis in the proton-rich region, we choose trajectory (e), which has the same entropy as trajectory (a) but a higher electron fraction of $0.55$.
Trajectories (f) and (g) have the same $Y_{e,9}$ as trajectory (e) but different $s$ ($129-141$).
We choose trajectory (h) to see the dependence of $Y_{e,9}$.
As an extreme case, we choose trajectory (i), which has the largest $\left<A\right>$ in the physically relevant proton-rich region.

In order to see gross features of the heavy-element synthesis, we define the average rates of individual reactions per nucleus in the heavy-element set ${\mathcal A_{>48}}$, defined in the previous subsection, as
\begin{eqnarray}
\lambda_{(p,\gamma)}&=&\frac{1}{Y_h}\sum_{i\in{\mathcal A_{>48}}}\rho N_{\rm A}\left<\sigma_{(p,\gamma)}(Z_i,A_i) v\right> Y_p Y(Z_i,A_i),\label{eq:pg}\\
\lambda_{(\gamma,p)}&=&\frac{1}{Y_h}\sum_{i\in{\mathcal A_{>48}}}\lambda_{(\gamma,p)}(Z_i+1,A_i+1) Y(Z_i+1,A_i+1),\nonumber\\\\
\lambda_{(n,\gamma)}&=&\frac{1}{Y_h}\sum_{i\in{\mathcal A_{>48}}}\rho N_{\rm A}\left<\sigma_{(n,\gamma)}(Z_i,A_i) v\right> Y_n Y(Z_i,A_i),\\
\lambda_{(\gamma,n)}&=&\frac{1}{Y_h}\sum_{i\in{\mathcal A_{>48}}}\lambda_{(\gamma,n)}(Z_i,A_i+1) Y(Z_i,A_i+1),\\
\lambda_{(n,p)}&=&\frac{1}{Y_h}\sum_{i\in{\mathcal A_{>48}}}\rho N_{\rm A}\left<\sigma_{(n,p)}(Z_i,A_i) v\right> Y_n Y(Z_i,A_i),
\end{eqnarray}
\begin{eqnarray}
\lambda_{(p,n)}&=&\frac{1}{Y_h}\sum_{i\in{\mathcal A_{>48}}}\rho N_{\rm A}\left<\sigma_{(p,n)}(Z_i-1,A_i) v\right> Y_p Y(Z_i-1,A_i),\nonumber\\\\
\lambda_{\beta^+} &=&\frac{1}{Y_h}\sum_{i\in{\mathcal A_{>48}}}\lambda_{\beta^+}(Z_i,A_i) Y(Z_i,A_i),\\
\lambda_{\beta^-} &=&\frac{1}{Y_h}\sum_{i\in{\mathcal A_{>48}}}\lambda_{\beta^-}(Z_i,A_i) Y(Z_i,A_i).\label{eq:beta-}
\end{eqnarray}
Here $\sigma_{(p,\gamma)}(Z,A)$, $\sigma_{(n,\gamma)}(Z,A)$, $\sigma_{(p,n)}(Z,A)$, and $\sigma_{(n,p)}(Z,A)$ are the cross sections of $(p,\gamma)$, $(n,\gamma)$, $(p,n)$, and $(n,p)$ reactions of nuclear species ${}^{A}Z$.
$\lambda_{(\gamma,p)}(Z,A)$ and $\lambda_{(\gamma,n)}(Z,A)$ are reaction rates of the photo-disintegrations $(\gamma,p)$ and $(\gamma,n)$.
$\lambda_{\beta^+}(Z,A)$ and $\lambda_{\beta^-}(Z,A)$ are $\beta^+$ decay and $\beta^-$ decay rates.
The rates defined in Equations (\ref{eq:pg})-(\ref{eq:beta-}) represent the reaction rates of dominant isotopes at a given time.
In addition, we define the average proton number $\left<Z\right>$ of the heavy-element set and the average neutron number $\left<N\right>$ of them as
\begin{eqnarray}
\left<Z\right> &=& \frac{1}{Y_h} \sum_{i\in{\mathcal A_{>48}}} Z_i Y(Z_i,A_i),\\
\left<N\right> &=& \frac{1}{Y_h} \sum_{i\in{\mathcal A_{>48}}} N_i Y(Z_i,A_i).
\end{eqnarray}

\subsubsection{Nucleosynthesis Result for the Fiducial Parameter Set {\rm (a)}}

First, we show the result of the nucleosynthesis of the fiducial wind trajectory with $\epsilon_{\nu_e}=11\ {\rm MeV}$ and $\epsilon_{\bar{\nu}_e}=16\ {\rm MeV}$, the parameter set of which is based on the result of \cite{2012PTEP.2012aA304S}.
This trajectory corresponds to point (a) in Figure \ref{fig:fiducial}.
The electron fraction $Y_{e,9}$ of this trajectory is 0.48.
The $r$-process occurs, and the average mass number becomes $\left<A\right>=118$ (see Table \ref{tab:trajs}).

\begin{figure}[htbp]
\begin{center}
\includegraphics[width=1\hsize]{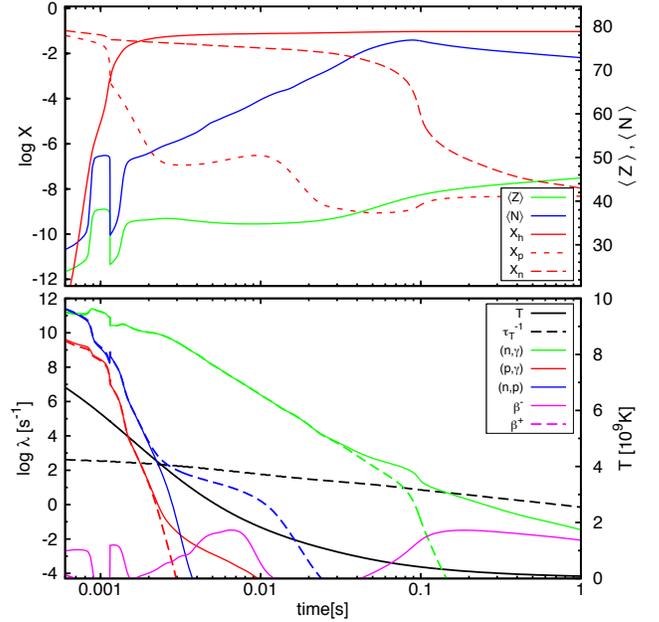}
\caption{
Top: time evolution of the mass fractions of protons, neutrons, and the total value of the heavy elements, and the average proton and neutron number of them in trajectory (a), which has the parameter set based on \cite{2012PTEP.2012aA304S}.
Bottom: time evolution of the average reaction rates per nucleus in the heavy-element set in the same trajectory.
The rates of $(n,\gamma)$, $(p,\gamma)$, and $(n,p)$ reactions are shown as green, red, and blue solid lines, respectively, and their inverse reaction rates are shown in dashed lines.
The rates of $\beta^-$-decays (solid line) and $\beta^+$-decays (dashed\ line) are shown as pink lines.
In addition to them, the temperature is shown as a black solid line, and the temperature-decrease rate $\tau_T^{-1}$ is shown as the black dashed line.
}
\label{fig:traj_a}
\end{center}
\end{figure}

\begin{figure}[htbp]
\includegraphics[width=\hsize]{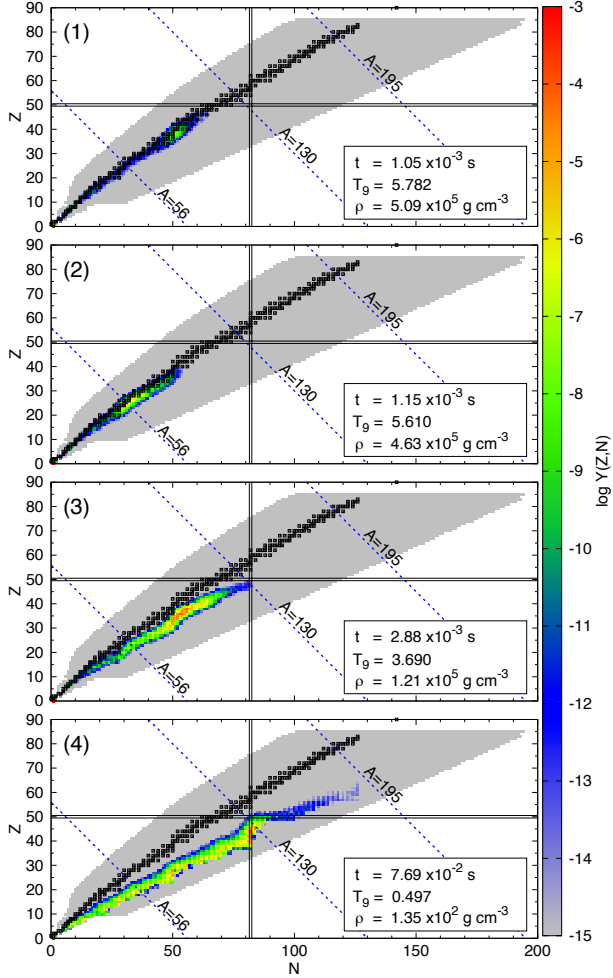}
\caption{
Snapshots of the nuclear abundance distribution of nucleosynthesis in the $N-Z$ plane for fiducial trajectory (a).
The blue dashed lines indicate the nuclei of $A=56$, 130, and 195.
Stable nuclei are shown by black squares.
Vertical and horizontal belts represent the $N=82$ neutron magic number and the $Z=50$ proton magic number, respectively.
}
\label{fig:traj_a_snaps}
\end{figure}

Figure \ref{fig:traj_a} shows the time evolution of the mass fractions of neutrons, protons, and the heavy-element set (top panel) and the average reaction rates of the heavy-element set ($\lambda_{(p,\gamma)}$, $\lambda_{(\gamma,p)}$, $\lambda_{(n,\gamma)}$, $\lambda_{(\gamma,n)}$, $\lambda_{(n,p)}$, $\lambda_{(p,n)}$, $\lambda_{\beta^-}$, and $\lambda_{\beta^+}$; bottom panel).
We show the snapshots of the abundance distribution in Figure \ref{fig:traj_a_snaps}.

We explain the time evolution of the nucleosynthesis in this trajectory.
Nucleons are still free particles at a temperature higher than $T_9\approx 6$.
After the temperature drops below $T_9 \approx 6$, the heavy elements up to $A\sim 100$ are temporarily produced, and the average charge number becomes $\left<Z\right> = 38$ (see top panel of Figure \ref{fig:traj_a} and \ref{fig:traj_a_snaps}-(1)).
This is because the numbers of nucleons are larger than that of the corresponding quasi-statistical equilibrium (QSE) value \citep{2002PhRvL..89w1101M}.
After this phase, the nuclei heavier than iron-group are photo-disintegrated into lighter nuclei as the free nucleons are recombined into alpha particles at $T_9\approx 5.5$.
The nuclei of $A\sim 80-90$ are not disintegrated completely in this phase because the nuclei at the neutron magic number at $N=50$ have relatively larger binding energies (see Figure \ref{fig:traj_a_snaps}-(2)).

The $r$-process begins after the temperature decreases below $T_9 \approx 4$.
In this phase, neutron capture reactions proceed until heavy nuclei reach the classical waiting points, where the reaction $(n,\gamma)$ balances with $(\gamma,n)$ (see top panel of Figure \ref{fig:traj_a} and Figure \ref{fig:traj_a_snaps}-(3)).
In the ordinary $r$-process, the nuclei on the waiting points can obtain additional neutrons through the $\beta^-$ decay.
In the case here, however, the $(p,n)$ reactions play the role.
The protons are supplied through the electron neutrino absorption reaction of free neutrons ($n+\nu_e\rightarrow p+e^-$).

At the time $t\sim0.01\ {\rm s}$, the rate of $(p,n)$ reactions decreases because of the low temperature and the heavy-element production is temporarily frozen at the $N=82$ magic nuclei.
After this phase, as the temperature decreases further, the rate of $(\gamma,n)$ reactions becomes too small to balance with $(n,\gamma)$, so the neutron capture reactions produce the heavy elements beyond $N=82$.
In this phase, the so-called cold $r$-process is realized, in which $(n,\gamma)$ reactions and $\beta^-$ decay reactions have main contributions to produce heavy nuclei \citep{2007ApJ...666L..77W}.
The nuclear abundance distribution, as functions of neutron and proton numbers, at $T_9\approx 0.5$ is shown in Figure \ref{fig:traj_a_snaps}-(4).
In this figure, we see that, although the fraction is very small, nuclei of $N>82$ are produced by the $r$-process.
After the phase ($T_9 \lesssim 0.5$), all reaction rates get small and the abundance distribution is completely frozen.

\begin{figure*}[t]
\begin{minipage}[t]{0.47\hsize}
   \includegraphics[width=\hsize]{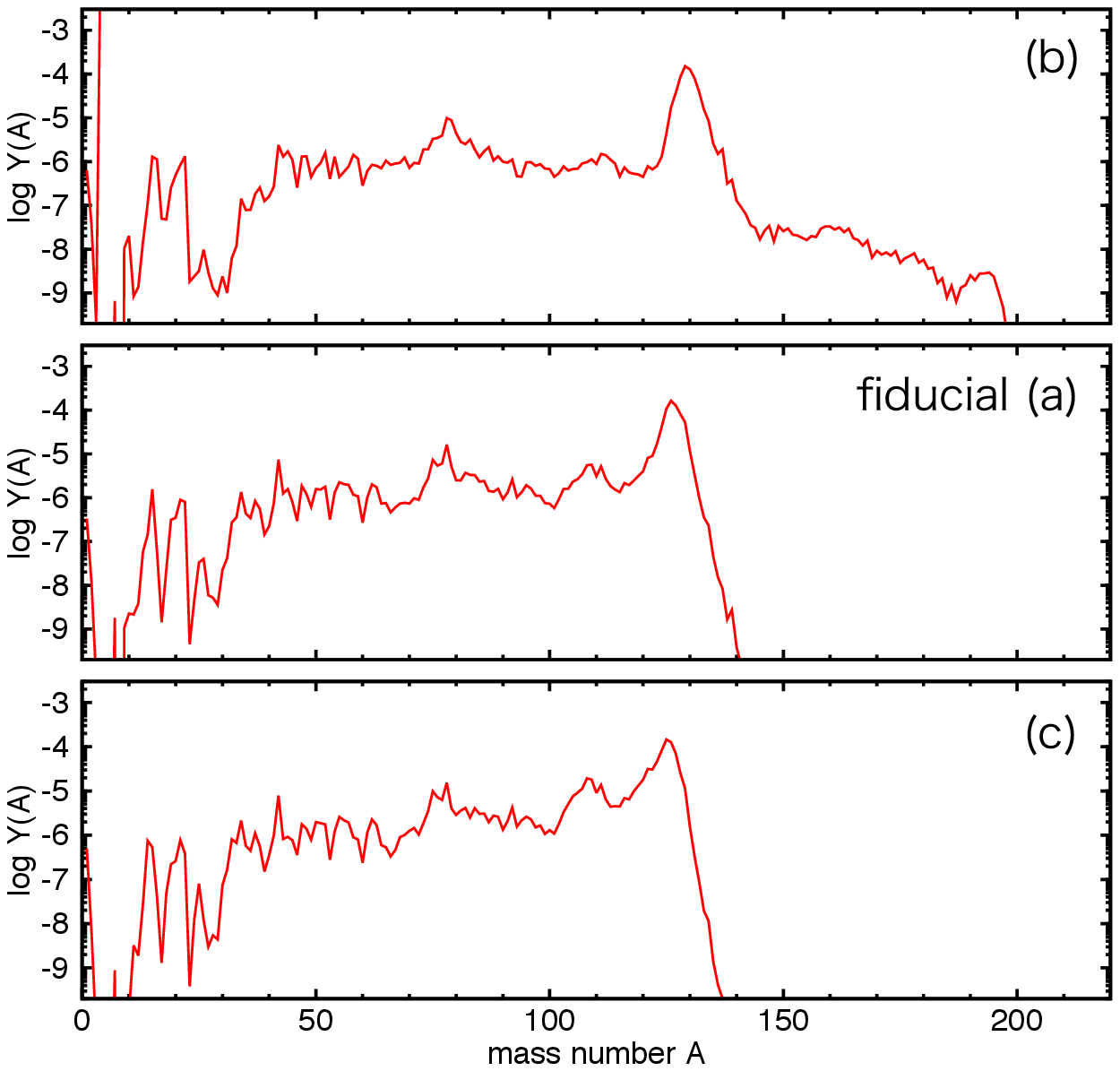}
\caption{
Abundance distributions of trajectories (b) (top), (a) (middle), and (c) (bottom).
These trajectories have almost the same electron fractions and different entropies (see Table \ref{tab:trajs}).
}
 \label{fig:abundance-abc}
 \end{minipage}
 \hspace{7mm}
 \begin{minipage}[t]{0.47\hsize}
    \includegraphics[width=\hsize]{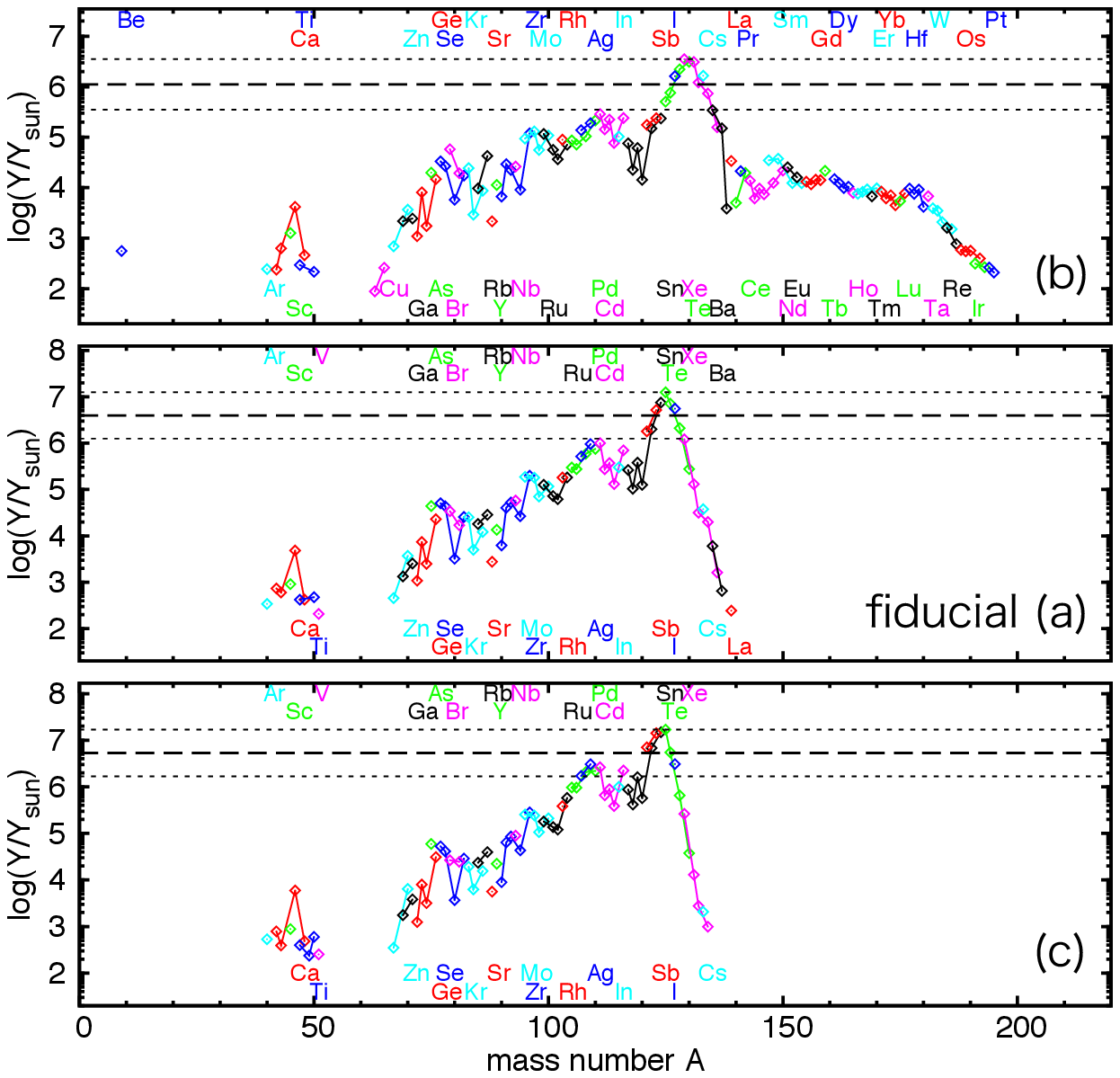}
 \caption{
Production factors of the nuclei obtained from the nucleosynthesis in trajectory (b) (top), (a) (middle), and (c) (bottom).
The solid line connects isotopes of a given element.
The dotted horizontal lines indicate the largest production factor and a factor of 10 smaller than that.
The dashed horizontal line indicates the median value.
 }
 \label{fig:prdfctr-abc}
 \end{minipage}
\end{figure*}

The middle panel of Figure \ref{fig:abundance-abc} shows the abundance distribution of trajectory (a).
The nuclei in the second peak elements are mainly produced.
The mass number of the most abundant nucleus in the heavy element is $A=126$.

The middle panel of Figure \ref{fig:prdfctr-abc} shows the production factors of final abundances of the nucleosynthesis result of trajectory (a).
The production factor is defined as the abundance ratio to the solar system composition \citep{1989GeCoA..53..197A}.
It is seen that the nuclei that form the second-peak are mainly produced (see mainly produced nuclides in Table \ref{tab:trajs}).

\subsubsection{Parameter Sets in Neutron-rich Region {\rm (}b{\rm )}-{\rm (}d{\rm )}}

Next, we show the dependence of the abundance distribution on the neutrino parameters in the trajectories in the neutron-rich region $Y_{e,9}<0.5$ ((b)-(d)), where the $r$-process occurs as in the fiducial trajectory (a).

The top and bottom panels of Figures \ref{fig:abundance-abc} and \ref{fig:prdfctr-abc} show the abundance distributions and the production factors of trajectories (b) and (c), which have the same electron fractions as the fiducial trajectory (a) ($0.48$) and different entropies ($138$ and $130$ for trajectory (b) and (c), respectively).
We see that the peak of the abundance distribution of trajectory (b) is $A=129$, which is slightly larger than that of the fiducial trajectory (a) ($A=126$), and the second-peak elements such as Te, I, Xe, and Cs are produced (see Table \ref{tab:trajs} for individual isotopes).

In trajectory (c), on the other hand, the abundance peak is $A=125$, which is almost the same as trajectory (a), and mainly produced elements are also the same (i.e., the isotopes of Sn, Sb, Te, and I).
In addition to these elements, lighter elements such as Ag, Cd, and Pd are co-produced.
This is because the $r$-process does not proceed sufficiently and the nuclei that do not pass through the $N=82$ magic number finally decay into these elements.

The abundance distribution of trajectory (b) is slightly shifted to heavier-element production, and the distribution of trajectory (c) indicates an opposite trend, but their difference is small.

\begin{figure}[htbp]
\includegraphics[width=\hsize]{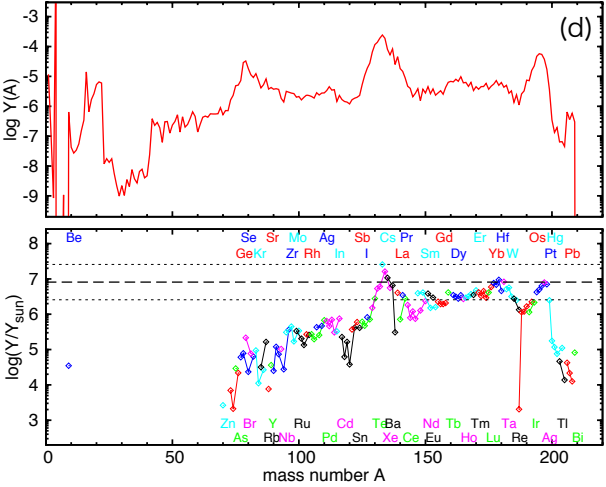}
\caption{Final abundance distribution (top) and the production factor (bottom) of trajectory (d).}
\label{fig:traj_d}
\end{figure}

We will show the nucleosynthesis result of a more neutron-rich trajectory with the entropy similar to trajectory (a).
The electron fraction of trajectory (d) is $Y_{e,9}= 0.42$ (see Table \ref{tab:trajs}).
Figure \ref{fig:traj_d} shows the abundance distribution and the production factors of the trajectory.
We clearly see the second peak and third peak in the distribution.
This means that a stronger $r$-process occurs and the nuclei of wide mass number range from second peak to the third peak are produced.
The time evolution of the average reaction rates is almost the same as in trajectory (a) but the neutron-to-seed ratio at the beginning of the cold $r$-process is about 30.
Therefore, we conclude that the nucleosynthesis results do not depend strongly on the entropy, but mainly on the electron fraction $Y_{e,9}$.

The combination $s/Y_{e,9} \tau_{T,0.5}^{1/3}$ is a useful quantity for estimating the neutron-to-seed ratio and whether the $r$-process produces the third-peak elements.
\cite{1997ApJ...482..951H} gave an approximate condition for making the third-peak ($A\sim 200$) elements in the trajectory as
\begin{equation}
f_{200} \equiv \frac{(s/230)}{(Y_{e,9}/0.4)(\tau_{T,0.5}/20\ {\rm ms})^{1/3}} \gtrsim 1. \label{eq:f200}
\end{equation}
\citep[This definition of $f_{200}$ is based on Equation (1) in][.]{2013ApJ...770L..22W}
Note that the condition is somewhat relaxed for the wind where $Y_e>0.46$ \citep[see Figure 10 in][]{1997ApJ...482..951H} because the rate of seed element production decreases in such a high-$Y_e$ circumstance.

The values of $f_{200}$ are shown in Table \ref{tab:trajs}.
It is about 0.696 in trajectory (a), which is below unity.
As a result, the neutron-to-seed ratio is not large, 10-20 at the beginning of the cold $r$-process, so the $r$-process in this trajectory is not strong enough to produce the third-peak nuclei.
On the other hand, the values of $f_{200}$ are 0.730 and 0.676 for trajectories (b) and (c), respectively, so the neutron-to-seed ratio is bigger in trajectory (b) and smaller in trajectory (c).
Therefore, slightly heavier and less heavier nuclei are synthesized in trajectories (b) and (c), respectively, as shown in Figure \ref{fig:abundance-abc}.

In trajectory (d), $f_{200}=0.795$, which is larger than those of the above trajectories because of the smaller value of $Y_e$.
Therefore, a stronger $r$-process proceeds and, as shown in Figure \ref{fig:traj_d}, the third-peak elements are produced in the trajectory.
(The third peak appears while the value $f_{200}$ is still smaller than the unity. This is because Equation (\ref{eq:f200}) is only  an approximate condition.)

\subsubsection{Parameter Sets in Proton-rich Region {\rm (}e{\rm )}-{\rm (}i{\rm )}}

The parameters suggested in \cite{2012PTEP.2012aA304S} are close to the contour of $Y_{e,9}=0.5$, so the initial electron fraction $Y_{e,9}$ of the trajectories in a certain range of the parameter values $\epsilon_{\nu_e}$ and $\epsilon_{\bar{\nu}_e}$ becomes greater than 0.5.
In these trajectories, the $\nu p$-process occurs.
Here we explore the nucleosynthesis and the abundance distributions of some proton-rich trajectories (e)-(i).

\begin{figure}[htbp]
   \includegraphics[width=\hsize]{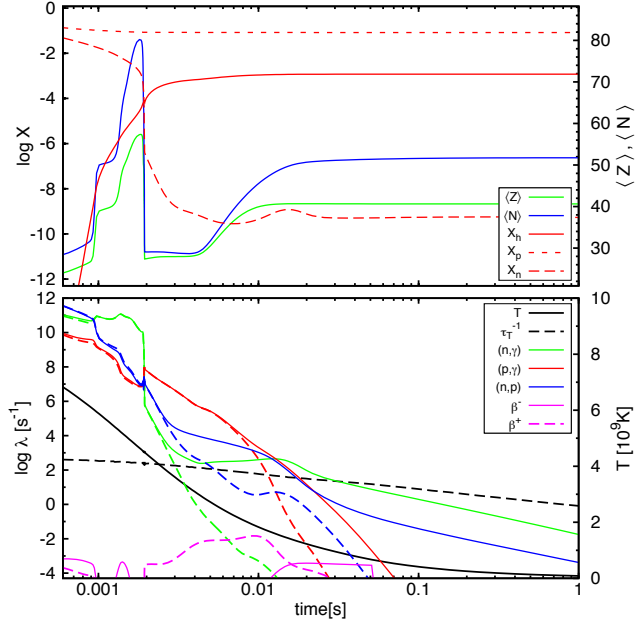}
\caption{Same figure as Figure \ref{fig:traj_a}, but for the result of trajectory (e).}
\label{fig:traj_e}
\end{figure}

\begin{figure}[htbp]
\includegraphics[width=\hsize]{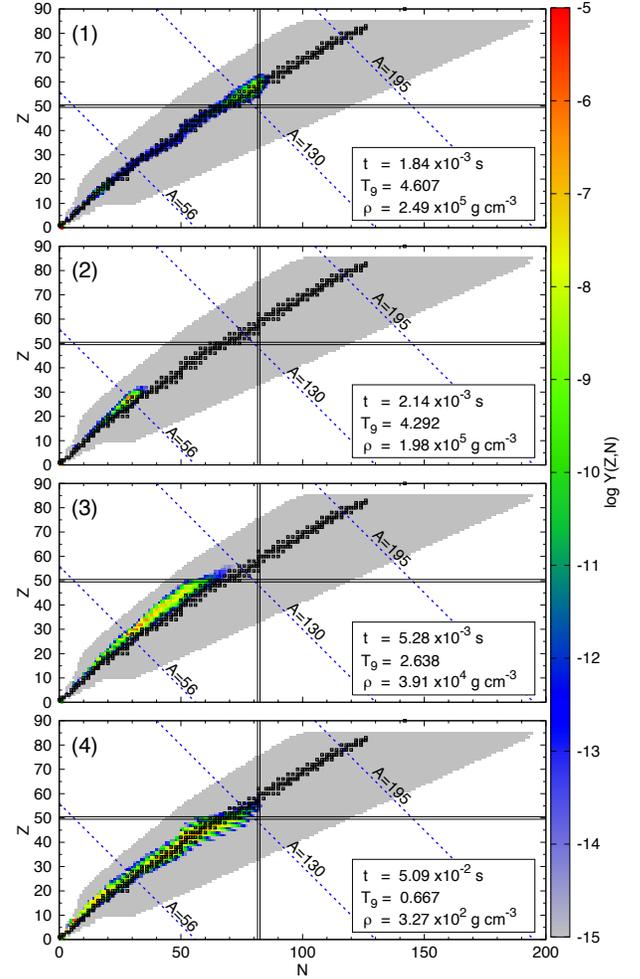}
\caption{
Snapshots of the nuclear abundance distribution of nucleosynthesis for the trajectory (e).
}
\label{fig:traj_e_snaps}
\end{figure}

Figure \ref{fig:traj_e} shows the time evolution of the nucleosynthesis and the abundance distribution of trajectory (e), of which $Y_{e,9}$ and $s$ are 0.55 and 133, respectively.
Initially, the material is composed mainly of free protons and neutrons.
Then, heavy nuclei are temporarily produced as in the neutron-rich case (see top panel of Figures \ref{fig:traj_e} and \ref{fig:traj_e_snaps}-(1)).

At temperature $T_9\approx 4.3$, the heavy nuclei are completely photo-disintegrated into iron-group nuclei and the mass fraction of neutron $X_n$ decreases to $\sim 10^{-9}$ (see top panel of Figures \ref{fig:traj_e} and \ref{fig:traj_e_snaps}-(2)).
After this phase, neutrons are supplied through the electron antineutrino absorption on free protons ($p+\bar{\nu}_e\rightarrow n+e^+$).
After the temperature decreases to $T_9\approx 4$, the balance between $(n,p)$ and $(p,n)$ reactions is broken, and then the $\nu p$-process starts to operate.
In this phase, $(p,\gamma)$ and $(\gamma,p)$ reactions are balancing, and most of the heavy nuclei are in the waiting point of the $\nu p$-process.
The production of the heavy elements occurs via $(n,p)$ reactions.
The path of the $\nu p$-process ends at $Z=50$ proton magic nuclei.
The reaction flow hardly passes through $Z=50$ because of small cross sections of $(p,\gamma)$ reactions of $Z=50$ isotopes (see Figure \ref{fig:traj_e_snaps}-(3)).
When the temperature gets below $T_9\approx 2.5$, the balance between $(\gamma,p)$ and $(p,\gamma)$ reactions in broken.
In this phase, $(n,p)$ reactions gradually take over from $(\gamma,p)$ reactions since neutrons do not feel the Coulomb barrier.

When the temperature decreases below $T_9\approx 1.5$, the heavy nuclei are concentrated at the stable line and $(n,\gamma)$ reaction rates become larger than $(n,p)$ reaction rates.
As a result, a fraction of large $A$ nuclei become neutron-rich (see Figure \ref{fig:traj_e_snaps}-(4)).
After this phase, all reaction rates get small and the abundances are completely frozen.

\begin{figure*}[t]
\begin{minipage}[t]{0.47\hsize}
   \includegraphics[width=\hsize]{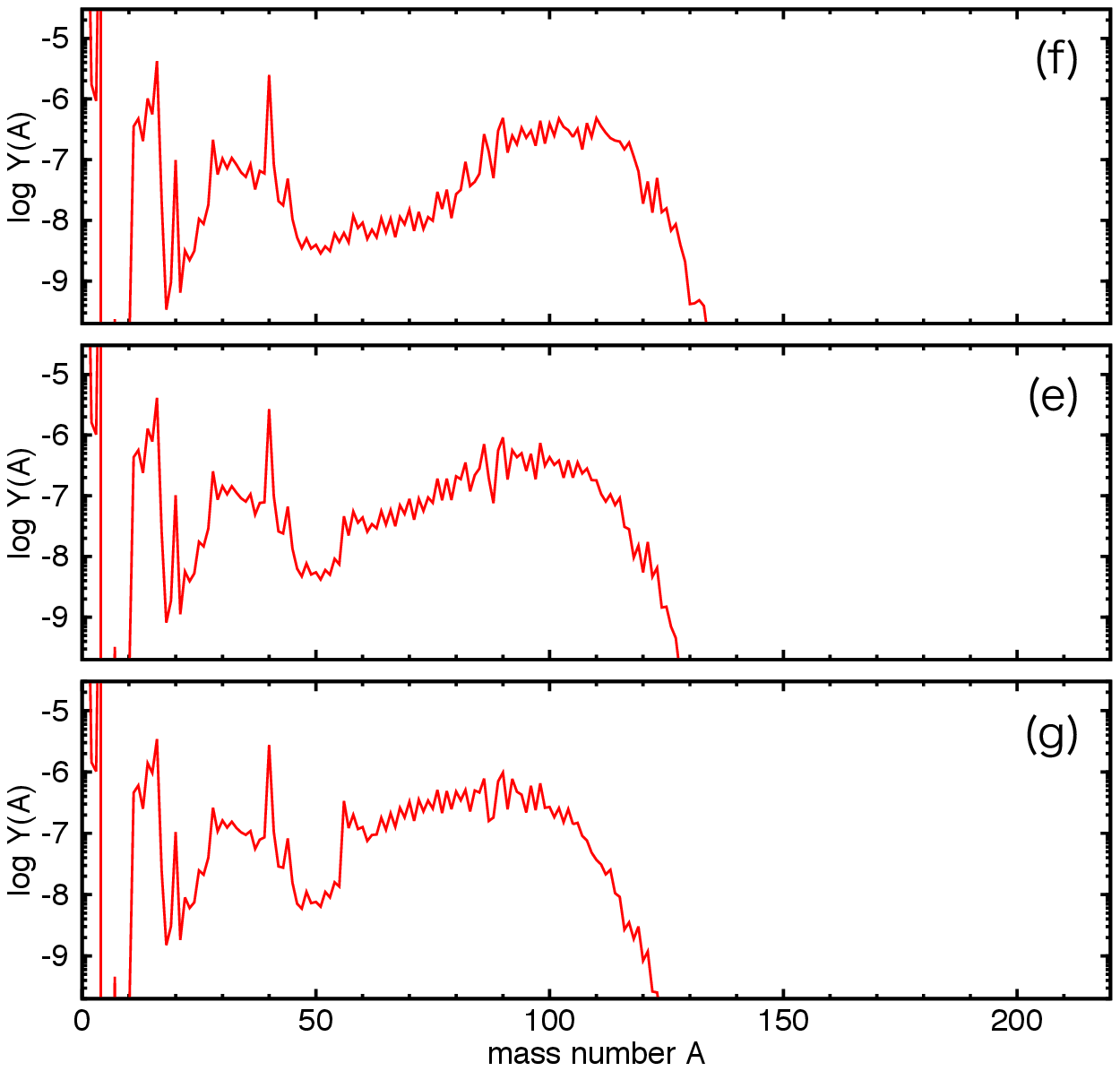}
\caption{
Same as Figure \ref{fig:abundance-abc}, but for the proton-rich trajectories (f) (top), (e) (middle), and (g) (bottom).
}
\label{fig:abundance-efg}
\end{minipage}
\hspace{7mm}
\begin{minipage}[t]{0.47\hsize}
    \includegraphics[width=\hsize]{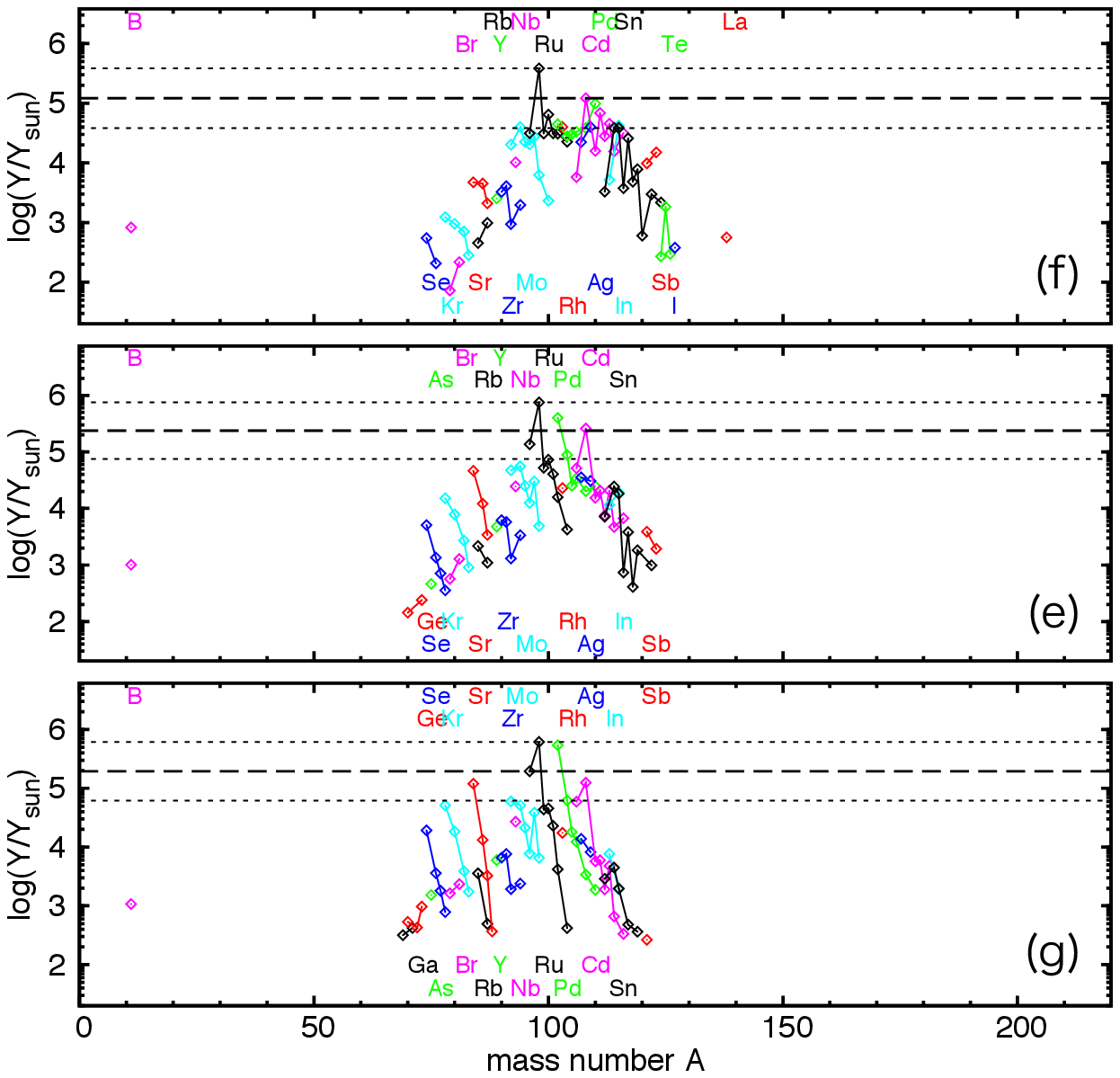}
\caption{
Same as Figure \ref{fig:prdfctr-abc}, but for the proton-rich trajectories (f) (top), (e) (middle), and (g) (bottom).
}
\label{fig:prdfctr-efg}
\end{minipage}
\end{figure*}

The abundance distribution of trajectory (e) (middle panel of Figure \ref{fig:abundance-efg}) indicates a broad peak of $A\sim 60-120$.
The middle panel of Figure \ref{fig:prdfctr-efg} shows the production factors of final abundances of this trajectory.
It is seen that $A\sim100$ proton-rich isotopes, such as Ru, Pd, and Cd, are produced.

We also consider trajectories (f) and (g), of which the electron fraction is almost the same as for trajectory (e) but the entropy is different ($141$ and $129$ for trajectories (f) and (g), respectively).
The top and bottom panels of Figure \ref{fig:abundance-efg} show the abundance distributions of these trajectories.
The $\nu p$-process scarcely passes through the $Z=50$ magic number in either of the trajectories, so the peaks of their abundance distributions are almost unchanged at $A\sim100$.
The top and bottom panels of Figure \ref{fig:prdfctr-efg} show the production factors of these trajectories.
They are almost the same as that of trajectory (e) except that the peak of the production factors slightly shifts to a larger mass number in trajectory (f).

In trajectories that have larger $Y_{e,9}$, heavier nuclei are produced compared with trajectory (e).
In trajectory (h), which has the electron fraction $Y_{e,9}=0.57$, for example, the isotopes of Ru, Pd, Cd, In, and Sn are abundantly produced (see Table \ref{tab:trajs} for details).
However, the produced elements that have $A\gtrsim 100$ are not proton-rich because of efficient $(n,\gamma)$ reactions in the phase of $T_9\lesssim 2$.

\begin{figure}[t]
\includegraphics[width=1.1\hsize]{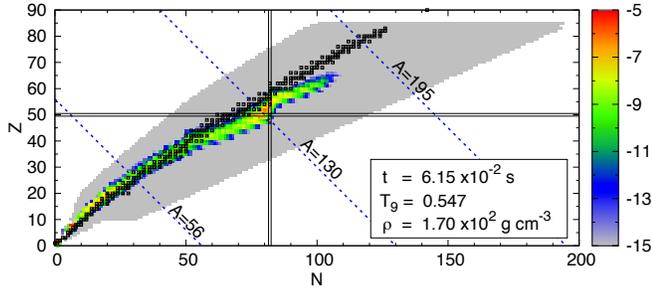}
\caption{
Nuclear abundance distribution at $T_9\approx0.55$ in trajectory (i).
}
\label{fig:ab-traj_i}
\end{figure}

\begin{figure}[htbp]
   \includegraphics[width=\hsize]{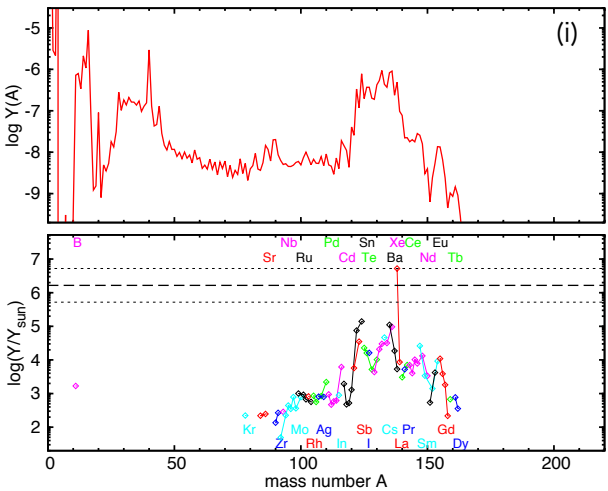}
\caption{Final abundance distribution (top) and production factor (bottom) of trajectory (i).}
\label{fig:traj_i}
\end{figure}

In trajectory (i), the most extreme (but not unphysical) trajectory in Figure \ref{fig:fiducial}, efficient $(p,\gamma)$ reactions due to a large number of free protons proceed strong $\nu p$-process nucleosynthesis and heavy nuclei can be produced beyond the $Z=50$ magic number.
Figure \ref{fig:ab-traj_i} shows the nuclear abundance distribution at $T_9\approx 0.55$ in trajectory (i).
It is seen that a large amount of nuclei pass through $Z=50$ and are concentrated in a neutron-rich region.

In Table \ref{tab:trajs}, we show the quantity $\Delta_n$, the number of neutrons per seed nucleus produced via $p+\bar{\nu}_e\rightarrow n+e^+$ reaction for $T_9<3$ \citep[see Equation (1) in ][]{2011ApJ...729...46W}.
When $\Delta_n$ is larger, more $(n,p)$ reactions occur during the $\nu p$-process so that heavier nuclei are produced.
In higher-entropy trajectories, less seed nuclei are produced and $\Delta_n$ is larger.
In trajectory (i), $\Delta_n=110$, which is much larger than that of (e)-(g) (39.7, 49.2, and 31.1, respectively).
This is the reason why the $\nu p$-process can pass through the $Z=50$ magic number and the nuclei are concentrated in a neutron-rich region in this trajectory.

\cite{2011ApJ...729...46W}, a comprehensive study of the $\nu p$-process in the ordinary PNS wind,
showed that heavy $p$-nuclei of $A\gtrsim110$ can be synthesized through the $\nu p$-process
only in extreme conditions where $Y_{e}>0.6$.
In our HN wind model, however, the $\nu p$-process produces heavy nuclei of $A\gtrsim100$ for smaller values of $Y_e$, although these are not $p$-nuclei.
This is because the higher neutrino luminosity causes less seed element production and more efficient neutron production through $p+\bar{\nu}_e\rightarrow n+e^+$ reactions.
As a result, the value of $\Delta_n$ can become larger than 100, which is not achieved even if $Y_e$ is set as 0.8 in the ordinary PNS wind \citep{2011ApJ...729...46W}.
Heavy $p$-nuclei of $A=110-125$ are underproduced in the $\gamma$-process \citep{2002ApJ...576..323R}, so our HN wind may be a possible site for this $p$-nuclei synthesis if produced nuclei are $p$-nuclei.
This may be possible when we consider the wind termination shock as indicated in Section \ref{sec:discussion}.

Only a small amount of nuclei in the neutron-rich region go beyond the $N=82$ magic number during the nucleosynthesis.
Thus, the $A\sim 130-140$ peak is produced in the final abundance distribution, as shown in the top panel of Figure \ref{fig:traj_i}.
The bottom panel of Figure \ref{fig:traj_i} shows the production factors of this trajectory.
The figure shows that ${}^{138}{\rm La}$ is mainly produced.

\section{Discussion}
\label{sec:discussion}

\subsection{The Effect of Heavy-element Synthesis on the Galactic Chemical Evolution}

In previous sections, we investigated the heavy-element synthesis in the neutrino-driven wind from mPNSs in the HN scenario.
It is suggested that HNe may have contributed to the synthesis of heavy elements and to the Galactic chemical evolution in the early universe because HNe may have occurred in Population~${\rm I\hspace{-.1em}I\hspace{-.1em}I}$ (Pop~${\rm I\hspace{-.1em}I\hspace{-.1em}I}$) stars \citep[e.g.,][]{2005ApJ...619..427U,2007ApJ...657L..77T,2007ApJ...660..516T}.
Therefore, here we discuss the effect of heavy-element synthesis in the mPNS wind on the Galactic chemical evolution.

It is known, from observations, that a large amount of ${}^{56}{\rm Ni}$ is produced in HNe.
Therefore, we discuss the effect of HNe on the Galactic chemical evolution by comparing the production factors of heavy elements produced in the neutrino-driven wind with that of ${}^{56}{\rm Ni}$ (Fe) produced in HNe.
If the production factor of a heavy element is larger than or equal to that of Fe, it can be said that the heavy-element synthesis in the neutrino-driven wind of HNe affects the Galactic chemical evolution.

We will evaluate the total production factors of elements considering the total ejecta of HNe.
The total production factor of Fe can be evaluated from observational analyses of the ${}^{56}{\rm Ni}$ yield in recent HNe.
The total production factor of Fe, $P_{\rm tot}({\rm Fe})$, is evaluated as
\begin{eqnarray}
P_{\rm tot}({\rm Fe}) = \frac{M({}^{56}{\rm Ni})}{X_\odot({\rm Fe})M_{\rm ini}},
\end{eqnarray}
where $M({}^{56}{\rm Ni})$ is the ${}^{56}{\rm Ni}$ yield, $X_\odot({\rm Fe})$ is the solar mass fraction of Fe, which is $\approx 1\times 10^{-3}$, and $M_{\rm ini}$ is the initial mass of the HN progenitor.
Table \ref{tab:HNe} shows the properties of observed HNe and the total production factor of Fe.
We see from the table that the total production factor of Fe $P_{\rm tot}({\rm Fe})$ is about $\gtrsim4$.
\begin{table*}[t]
\caption{
Properties of Observed Hypernovae
}
\centering
\begin{tabular*}{\hsize}{@{\extracolsep{\fill}}lcccccc}
\hline \hline
Name & Type & $E_{51}$ & $M_{\rm ini}/M_\odot$ & $M({}^{56}{\rm Ni})/M_\odot$ & $P_{\rm tot}({\rm Fe})$ & Reference \\ \hline
SN 1997ef & Ic & $8$ & $ 30-35$ & 0.15 & $4-5$ & \cite{2000ApJ...534..660I}\\
SN 1998bw & Ic & $30$ & $ 40$ & 0.7 & $17.5$ & \cite{1998Natur.395..672I} \\
SN 2002ap & Ic & $7$ & $21$ & 0.10 & $4$ & \cite{2002ApJ...572L..61M}\\
SN 2003dh & Ic & $30-50$ & $35-40$ & 0.35 & $8-9$ & \cite{2003ApJ...599L..95M} \\
SN 1999as & Ic & $20-50$ & $60-80$ & $>4$ & $>42-60$ & \cite{2001ASPC..251..238D}\\ 
SN 2003lw & Ic & $60$ & $40-50$ & 0.55 & $11-14$ & \cite{2006ApJ...645.1323M}\\ \hline
\end{tabular*}
\begin{tablenotes}
\scriptsize
\item {\bf Note.} $E_{51}$ is the explosion energy of SN in units of $10^{51}$ erg.
\end{tablenotes}
\label{tab:HNe}
\end{table*}
Thus, the total production factor of the element that can contribute the Galactic chemical evolution should be greater than $\sim10$ at least.
The total production factors of the heavy elements produced in the wind are estimated as
\begin{eqnarray}
P_{\rm tot}({}^{A}Z)&=&\frac{M_{\rm ej,wind}}{M_{\rm ej,tot}}P_{\rm wind}({}^{A}{Z})\nonumber \\
&\approx& \frac{\dot{M}_{\rm wind} \tau_{\rm NS}}{M_{\rm ej,tot}}P_{\rm wind}({}^{A}{Z})\nonumber \\
&\approx& 10^{-4}P_{\rm wind}({}^{A}{Z}) \Biggl(\frac{\dot{M}_{\rm wind}}{4\times10^{-3} M_\odot\ {\rm s^{-1}}}\Biggr)\nonumber\\
&&\hspace{10mm} \times \Biggl(\frac{\tau_{\rm NS}}{1\ {\rm s}}\Biggr)\Biggl(\frac{M_{\rm ej,tot}}{40M_\odot}\Biggr)^{-1},
\end{eqnarray}
where $P_{\rm wind}({}^{A}{Z})$ is the production factor of the element ${}^{A}Z$ in the wind ejecta, which is shown in Section \ref{sec:results}, $\dot{M}_{\rm wind}$ is the mass outflow rate of the wind, and $\tau_{\rm NS}$ is the lifetime of mPNSs.
Thus, the elements that have the production factor of winds $P_{\rm wind}({}^{A}{Z})\gtrsim10^{5}$ may contribute to the Galactic chemical evolution.

For the neutron-rich wind near the fiducial parameter set, as seen in Figure \ref{fig:prdfctr-abc}, the second-peak elements such as Sb, Sn, Te, and I are co-produced with these production factors larger than about $10^6$.
Therefore, the mPNS wind may contribute to the Galactic evolution of these elements.
In a more neutron-rich region, e.g., in trajectory (d), the wide range of nuclei from the second peak to the third peak have sufficient production factors to contribute to the Galactic chemical evolution as seen in Figure \ref{fig:traj_i}.

On the other hand, in the proton-rich wind, proton-rich isotopes of $A\sim100$, such as Ru, Pd, and Cd have sufficient production factors larger than about $10^5$ (see Figure \ref{fig:prdfctr-efg} for trajectories (e)-(g)).
In more proton-rich wind, e.g., trajectory (i), the production factor of ${}^{138}{\rm La}$ is $\sim10^7$, which is much larger compared with other elements produced in this trajectory as seen in Figure \ref{fig:traj_i}.

We see from Table \ref{tab:HNe} that the total production factor of Fe tends to be larger when the progenitor mass of the HN is larger (for SN 1999as, it is about 50 at least).
Thus, the threshold of the wind production factor for contribution to the Galactic chemical evolution may be one order larger than $\sim10^5$ for the HNe of more massive stars.

\subsection{Heavy-element Abundance in Very Metal-poor Stars}

Can we explain the abundance patterns of $r$-rich, very metal-poor stars by this mPNS wind model?
As seen in the previous sections, the so-called main $r$-process, which produces all $r$-process elements up to the third peak, hardly occurs in our parameter sets.
Therefore, it is very hard to explain the origin of very metal-poor stars having the ``universality" of the $r$-process \citep[][]{2003ApJ...591..936S,2008ARA&A..46..241S} using our mPNS wind model.

\begin{figure}[htbp]
\includegraphics[width=\hsize]{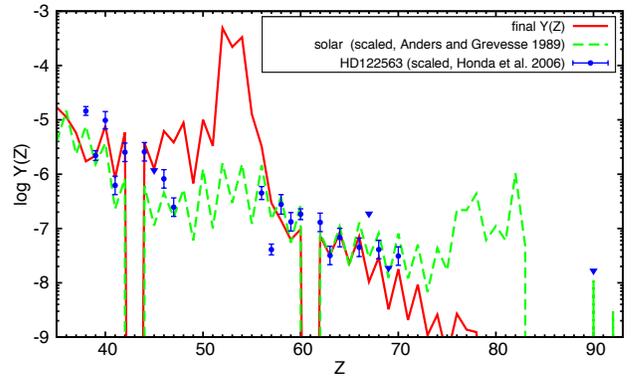}
\caption{
Abundance distribution of the elements as a function of atomic number $Z$.
The solid line indicates the nucleosynthesis result of the trajectory which has the parameter set $\epsilon_{\nu_e}=11.0\ {\rm MeV}$, $\epsilon_{\nu_e}=16.8\ {\rm MeV}$.
The dashed line is the scaled solar abundances \citep{1989GeCoA..53..197A}.
The points with error bars are the scaled abundances of HD~122563 \citep{2006ApJ...643.1180H}.
The inverted triangles show the upper limits of the elements.
The observational abundance patterns are scaled to the Eu abundance of the nucleosynthesis result.
}
\label{fig:obs}
\end{figure}

There are, however, some very metal poor stars that do not have that universality.
For example, HD~122563 is one of the stars, whose abundance pattern shows a gradually decreasing trend as a function of atomic number \citep{2006ApJ...643.1180H}.
They are called ``weak $r$ stars."
Thus, we compare our result with the abundance pattern of these weak $r$ stars.
Figure \ref{fig:obs} shows the abundance pattern of the nucleosynthesis result of the trajectory just near point (a) in Figure \ref{fig:fiducial}, which has the parameter set indicated in \cite{2012PTEP.2012aA304S}.
The parameter set of this trajectory is $\epsilon_{\nu_e}=11.0\ {\rm MeV}$, $\epsilon_{\nu_e}=16.8\ {\rm MeV}$.
In this figure, the scaled observational abundances of HD~122563 \citep{2006ApJ...643.1180H} are shown with points.
The abundance pattern shows an approximate agreement with the observational trend in $Z=39-45$ and $Z\gtrsim 58$.
Recent observations show that the heavy-element abundance pattern of weak $r$ stars has some varieties \citep[e.g.,][]{2007ApJ...666.1189H,2010ApJ...724..975R}.
These varieties may be explained by considering the difference of the properties of mPNS winds.

\subsection{The Dependence of Heavy-element Synthesis on the Mass, Radius, and Neutrino Luminosities of PNSs}
In this work, we investigated the dependence of heavy-element synthesis on the average energies of electron neutrinos and electron antineutrinos.
Here we discuss the dependence of the yields on the other wind parameters.

In addition to the uncertainties of average energies of electron neutrinos and electron antineutrinos $\epsilon_{\nu_e}, \epsilon_{\bar{\nu}_e}$, there will be uncertainties of the average energy of other flavors of neutrinos $\epsilon_{\nu_x}$ originated in approximated treatment of neutrino physics.
The neutral-current neutrino reactions with ${}^{4}{\rm He}$ will affect heavy-element synthesis.
However, this effect is less important than the charged-current neutrino reactions with $n$ and $p$.
Thus, this uncertainty will have only a minor impact on the heavy-element synthesis.
Luminosities of individual flavors of neutrinos $L_{\nu_e}$, $L_{\bar{\nu}_e}$, and $L_{\nu_x}$ also have uncertainties for the same reason.
These uncertainties are expected to affect the heavy-element synthesis through the change of the wind properties and the neutrino reactions.

On the other hand, the uncertainty of the radius of (the neutrinosphere of) the mPNS $R_\nu$ originates not only in the matter effect on the neutrino opacity but in the fact that the EOS of high-density nuclear matter, or the mass-radius relation of the NS, is still unknown.

The mass of a mPNS $M$ is another parameter that depends on the properties of its progenitor, such as initial mass, rotation, metallicity, and so on.
A recent study of Pop~${\rm I\hspace{-.1em}I\hspace{-.1em}I}$ star formation \citep{2013ApJ...773..185S,2014ApJ...781...60H} suggested that Pop~${\rm I\hspace{-.1em}I\hspace{-.1em}I}$ stars of the mass over $100M_\odot$ can be formed.
In the final stage of this evolution, these massive stars may have a massive iron core more than $3M_\odot$.
Recent observations suggest that the maximum mass of a neutron star is more than $2M_\odot$ \citep{2010Natur.467.1081D,2013Sci...340..448A}, so the EOS of the nuclear matter needs to be sufficiently stiff in order to support that mass.
In addition, Pop~${\rm I\hspace{-.1em}I\hspace{-.1em}I}$ stars cannot lose their angular momentum because of the absence of the mass loss.
Therefore, when these stars collapse, rapid rotation of these cores prevents them from collapsing to black holes, in addition to their thermal pressures.
Thus, the existence of mPNSs is quite possible when Pop~${\rm I\hspace{-.1em}I\hspace{-.1em}I}$ stars collapse.
In the studies about the nucleosynthesis of Pop~${\rm I\hspace{-.1em}I\hspace{-.1em}I}$ SNe and HNe \citep[e.g.,][]{2005ApJ...619..427U}, the ``mass cut," the remnant mass after the explosion, was set to explain the abundance patterns of the nuclei up to iron-peak elements of extremely metal-poor stars.
They changed the mass cut for different progenitor masses and explosion energies and set as $\approx3M_\odot$ in some cases.
Therefore, these studies imply that mPNSs may indeed be formed during Pop~${\rm I\hspace{-.1em}I\hspace{-.1em}I}$ HNe and the masses of them may change for different progenitor masses and explosion energies.
To discuss the history of heavy-element synthesis of HNe, further study is needed that covers a wide parameter range of the mass of HN progenitors.

Uncertainties and varieties of these parameters can affect the entropy $s$, the temperature-decrease timescale $\tau_T$, and the electron fraction $Y_e$ of the neutrino-driven wind and thus the nucleosynthesis processes.
In order to investigate the yields of the heavy elements of HNe, the systematic search of the nucleosynthesis changing these parameters is needed.
This also will appear in our next work.

In particular, we find an interesting nucleosynthesis feature in the parameter space.
There is a region where $\left<A\right>$ is large but the initial electron fraction $Y_{e,9}$ is only a little larger than 0.5, i.e., the region around $\epsilon_{\nu_e}\approx 9\ {\rm MeV}$ and $\epsilon_{\bar{\nu}_e}\approx 13\ {\rm MeV}$.
In the trajectories in this region, the abundances freeze out without the reaction equilibrium between the nucleons and alpha-particle.
As a result, after the temporal production of the $A\gtrsim100$ nuclei, photo-disintegration does not proceed completely, and nuclei around the $N=50$ magic number remain.
This process can occur in a realistic parameter range, but detailed analysis of the process in these trajectories is beyond the scope of this paper.
We will make a detailed analysis in our next work.

\subsection{Effect of Wind Termination}

As in the case of ordinary SNe, it could happen that the mPNS wind terminates 
at some radius because of the existence of the stellar material.
The wind termination can affect the nucleosynthesis because it 
changes the thermal histories of the fluid elements in the wind \citep{2008ApJ...672.1068K,2011A&A...526A.160A,2011ApJ...729...46W}.
Here we discuss the effect of the wind termination shock on the 
nucleosynthesis in the previous sections.

We assume that a termination shock appears at the radius $R_{\rm sh}$.
We apply the relativistic Rankine-Hugoniot jump conditions there.
The jump conditions are
\begin{eqnarray}
\rho_{\rm w} u_{\rm w}&=&\rho_{\rm s} u_{\rm s},\label{eq:jump1}\\
\rho_{\rm w}h_{\rm w} u_{\rm w}^2 + P_{\rm w}&=&\rho_{\rm s}h_{\rm s}u_{\rm s}^2 + P_{\rm s},\label{eq:jump2}\\
\rho_{\rm w}h_{\rm w}u_{\rm w}\sqrt{1+u_{\rm w}^2}&=&\rho_{\rm s}h_{\rm s} u_{\rm s}\sqrt{1+u_{\rm s}^2},\label{eq:jump3}
\end{eqnarray}
where $h=1+\epsilon+P/\rho$ is the specific enthalpy, and the subscripts 
``w" and ``s" indicate the quantity of the upper stream (wind) and 
downstream (shocked) material, respectively.
We assume that the velocity of the shock position is negligible compared 
to the wind velocity $\sim 0.3 c$.
The details of solving these equations are described in Appendix \ref{app:B}.
Solving the jump conditions (\ref{eq:jump1})-(\ref{eq:jump3}), we can 
evaluate the downstream quantities $u_{\rm s}$ and $T_{\rm s}$ from the upstream 
quantities $u_{\rm w}$ and $T_{\rm w}$ (the electron fraction is  assumed to 
be unchanged across the shock).
Then $u$, $T$, and $Y_e$ profiles in the shocked region ($r>R_{\rm sh}$) are 
obtained by Equations (\ref{eq:wind-mdot})-(\ref{eq:wind-qdot}) using the 
shocked values $u_{\rm s}$ and $T_{\rm s}$ as the boundary condition at $r=R_{\rm sh}$.

\subsubsection{The Effect on the $r$-process}

Here we discuss the effect of the wind termination shock on the 
nucleosynthesis result in the neutron-rich region in the parameter space.
A previous study investigated the effect of the wind termination shock 
assuming a shock radius of 500-30,000 km \citep{2008ApJ...672.1068K}.
They suggested that, in an ordinary SN, the wind termination jumps 
appear only when the temperature sufficiently decreases, so the wind 
termination has little effect on the final abundance pattern.
However, the wind termination shock in our HN wind may appear at 
sufficiently high temperature because of the high neutrino luminosity 
from the mPNS.
Thus, it can affect the final abundance pattern.

In the neutron-rich wind, when the post-shock temperature is high enough 
($T_9\gtrsim2\times10^9$K), the heavy elements can be photo-disintegrated 
into lighter nuclei.
Therefore, the wind termination shock can hinder the $r$-process.
\begin{figure*}[t]
\begin{minipage}[t]{0.47\hsize}
\includegraphics[width=\hsize]{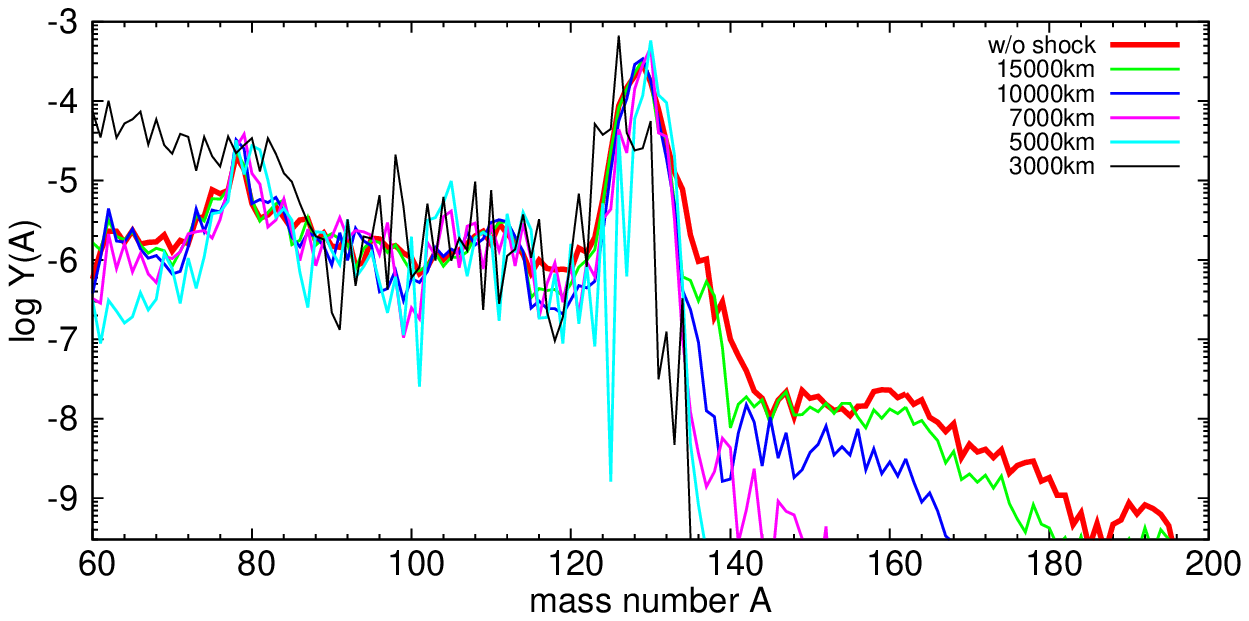}
\caption{
Abundance distribution of the elements as a function of mass number $A$.
The final abundance patterns of the trajectories of different shock 
radii (3000, 5000, 7000, 10,000, and 15,000 km) are shown.
The other parameters of the trajectories are the same as in trajectory (a), 
the result of which is also shown in the figure.
}
\label{fig:r-shock}
\end{minipage}
\hspace{7mm}
\begin{minipage}[t]{0.47\hsize}
\includegraphics[width=\hsize]{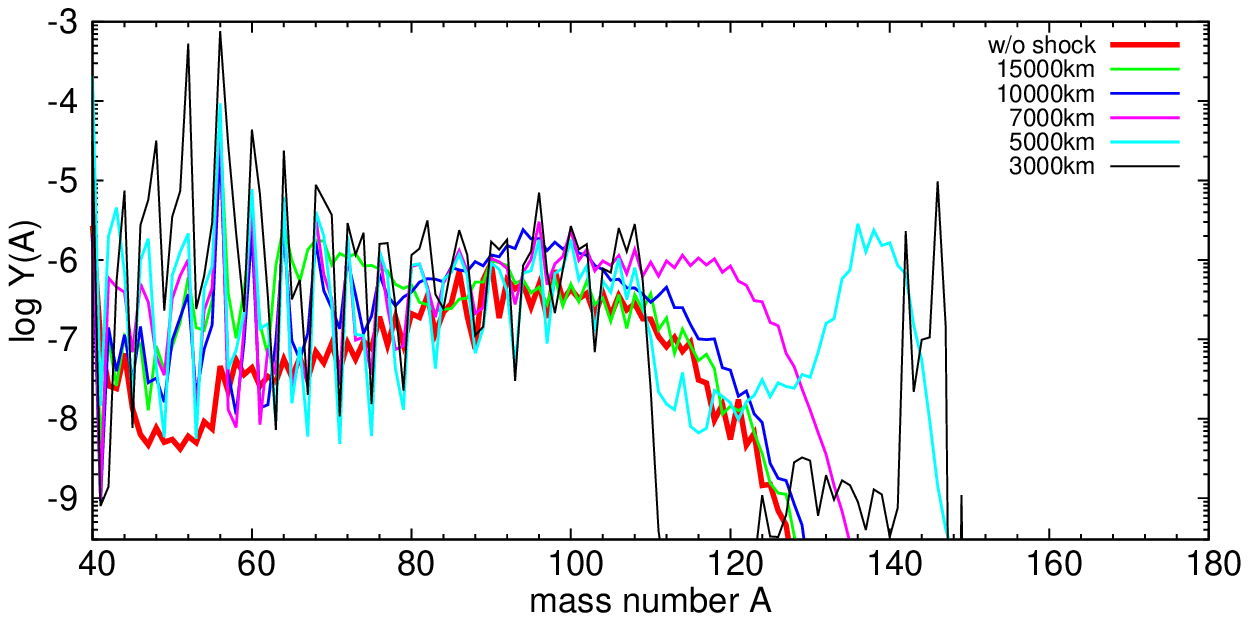}
\caption{
Same figure as Figure \ref{fig:r-shock}, but in the case of the 
proton-rich trajectories. The final abundance patterns of the 
trajectories of different shock radii (3000, 5000, 7000, 10,000, and 15,000~km).
Other parameters of the trajectories are the same as trajectory (e).
}
\label{fig:nup-shock}
\end{minipage}
\end{figure*}
In Figure \ref{fig:r-shock}, we show the nucleosynthesis result of 
the trajectories of different shock radii (3000, 5000, 
7000, 10,000, and 15,000 km).
The other parameters of the trajectories are the same as trajectory (a) 
(denoted as ``w/o shock" in Figure \ref{fig:r-shock}).
The abundances of the nuclei heavier than the second peak become 
smaller for smaller radius of the termination shock.
This is because, for a shock radius smaller than 10,000 km, 
the post-shock temperature increases to be $\gtrsim 1.5\times10^9$K 
and is sustained for a long time, and consequently, the elements larger 
than the second peak are photo-disintegrated into lighter nuclei.
In the case where the shock radius is larger than about 10,000 km, 
on the other hand, the post-shock temperature is below 
$1.5\times 10^9$K and the effect of the wind termination shock on 
the final abundance is small.

\subsubsection{The Effect on the $\nu p$-process}

The $\nu p$-process continues longer in the presence of the wind termination 
because the temperature decreases more slowly compared to the case without 
the wind termination.
As explained in \cite{2011ApJ...729...46W}, strong $\nu p$-processes
occur if the post-shock temperature becomes $T_9\gtrsim 1.5-3$.
Such a high post-shock temperature may be achieved in our HN wind model because of the higher neutrino luminosity.
Therefore, in such a case, the $\nu p$-process may proceed more efficiently than the wind without the termination shock.
From the viewpoint of heavy $p$-nuclei production, it is worth investigating 
the effect of the wind termination to study the synthesis of these elements in our HN wind model.

In Figure \ref{fig:nup-shock}, we compare the final abundance 
for the trajectories with different shock radii. 
First of all, note that the abundance pattern for $A\lesssim80$ is significantly
affected by the presence of the shock termination because proton-capture 
reactions in the region of smaller atomic number can proceed for a long time
owing to the higher temperature.

For the heavier region, on the other hand, the effect of the shock termination
depends on the location of the shock radius.
When the shock radius is smaller than 10,000 km, the post-shock temperature 
becomes high enough so that additional proton-capture reactions occur as expected above, and
consequently, the $\nu p$-process can produce heavier elements.
In addition, since the fluid elements are decelerated by the shock, 
they are exposed to large neutrino flux for a long time.
As a result, the $\nu p$-process proceeds for a long time.
Indeed, in the trajectories with the shock radius at small radii $R_{\rm sh}=5000$ 
and $7000$ km, the final abundance patterns are modified 
by the long-lasting $\nu p$-process.
In these trajectories, the heavy $p$-nuclei that are underproduced in the $\gamma$-process are indeed synthesized.
${}^{138}$La is produced in the trajectory of $R_{\rm sh}=5000$ km
and ${}^{113,115}$Sn are synthesized in the case of $R_{\rm sh}=7000$ km.
Thus, the HN wind model with the wind termination shock can be a possible production site of these $p$-nuclei.

More interestingly, in the case of $R_{\rm sh}=3000 {\rm km}$, the final abundance pattern 
totally changes from the result of the trajectory without the termination shock.
In this trajectory, the elements up to $A\sim150$ are once produced.
However, the elements of $A\sim110-140$ are photo-disintegrated 
because of the high post-shock temperature ($T_9\sim 2$).
The elements of the neutron magic number $N=82$ are relatively stable, 
so they avoid photo-disintegration.
As a result, the final abundance pattern has a sharp peak around $A = 145$.

Note that the final abundance pattern for the $\nu p$-process is less affected by the shock termination for the larger shock radius
because the temperature range relevant to the $\nu p$-process is higher that that for the $r$-process.

\subsection{The Effect of the Rotation and the Magnetic Field of the PNS}

In general, PNSs have an angular momentum and a magnetic field.
If the PNS has a strong magnetic field and rotates rapidly, the magnetic 
force is important for the dynamics of the wind and for the thermal 
history of the fluid element \citep{2007ApJ...659..561M,2014MNRAS.444.3537V}.
In \cite{2007ApJ...659..561M}, the criterion for which the wind is 
magnetically driven is shown (see their discussion above Equation (17) 
in their paper).
Using our parameter set (i.e., $L_{\nu,51} \sim 100$, $R_\nu=15$ km) and 
the quantities obtained our wind solutions  (the asymptotic velocity 
$v^a \sim 0.3c$ and the mass-loss rate $\dot{M}\sim 4\times 10^{-3}\ M_\odot\ {\rm s^{-1}}$), 
we can estimate the rotation period below which our HN wind is magnetically 
driven as

\begin{equation}
P \approx 0.6\ B_{14}\ {\rm ms},
\end{equation}
where $B_{14}$ is the magnetic field strength at the neutrinosphere 
in units of $10^{14}\ {\rm G}$. This is very close to the break-up 
period $\sim 0.6\ {\rm ms}$.
Therefore, the effect of the rotation and the magnetic field on 
the wind is negligible unless the rotation period is approximately the same as the 
break-up period or the magnetic field is strong enough.

If the short rotational period and the strong magnetic field are 
achieved, the temperature-decrease timescale can become even smaller. 
(It can be below 1~ms depending on the field strength and the 
rotational period as described in \cite{2007ApJ...659..561M}.)
However, the fluid element accelerated by the magnetic force can pass 
through the neutrino-heating region quickly.
Therefore, the magnetically driven wind has less asymptotic entropy 
than that of the wind driven only by neutrinos. It is important to 
investigate which of these opposite effects is dominant in our HN 
wind, but this is beyond the scope of this work.
We will investigate these effects next time.

\subsection{The Effect of Fallback}
The wind material could fall back to the central remnant (neutron star or black hole) with infalling stellar material.
Therefore, the ejected mass of our HN wind can be somewhat smaller than that estimated in the previous sections.

A numerical study about aspherical (jet-induced) explosion of a Pop III star \citep{2009ApJ...690..526T} suggests that if the energy deposition rate is high enough, the material just above the central remnant can be ejected toward the jet axis.
We consider the mass ejection toward the polar region so the wind material can avoid falling back because the ram pressure of the infalling stellar matter is expected to be weak in the region.

\section{Conclusions}

\label{sec:conclusions}
In this work, we investigated possible processes of heavy-element synthesis in HN explosions of very massive ($100M_\odot$) stars.
We considered the neutrino-driven winds from the mPNSs formed after the collapse.
We constructed a wind solution for the parameter set tabulated in Table \ref{tab:modelpar} based on the numerical relativity simulation \citep{2012PTEP.2012aA304S} and calculated nucleosynthesis on the winds.
We parameterized the average energies of electron neutrinos $\epsilon_{\nu_e}$ and electron antineutrinos $\epsilon_{\bar{\nu}_e}$ and investigated heavy-element synthesis processes.
Here we make our conclusions as follows.
\begin{enumerate}
\item The temperature-decrease timescale at $T=0.5\ {\rm MeV}$ of the mPNS wind (about $6-7$ ms) is much shorter than that of the ordinary PNS wind (about $40-50$ ms), and thus the environment of mPNS wind is conducive to heavy element synthesis.
\item In the neutron-rich region of the parameter space, although strong (i.e., making all elements up to third peak) $r$-process hardly occurs, weak $r$-process occurs and the second-peak elements are mainly synthesized.
Near the fiducial parameter, there is a parameter region in which the wind reproduces the weak $r$ elemental abundance pattern of some metal-poor stars, e.g., HD~122563.
The third-peak elements can be produced only in the narrow parameter region $\epsilon_{\bar{\nu}_e}-\epsilon_{\nu_e}\gtrsim 7\ {\rm MeV}$.
\item In the moderately proton-rich region, $\nu p$-process occurs and some light $p$-nuclei (e.g., proton-rich isotopes of Ru, Pd, and Cd) are synthesized.
In some parameter region around $\epsilon_{\bar{\nu}_e}\approx \epsilon_{\nu_e}$, strong $(n,\gamma)$ reactions proceed and make the abundance distribution into a more neutron-rich one.
As a result, the neutron-rich elements between $A=130-140$ are mainly produced.
\end{enumerate}
As discussed in the previous section, the uncertainties in the other parameters such as $L_{\nu_i}$, $M$, and $R_\nu$, which were unchanged in this work, are also expected to affect the nucleosynthesis in the wind from mPNSs.
In the next study, we will discuss the possibility of HNe to the Galactic chemical evolution through the parameter survey of the nucleosynthesis in the wind about these parameters.

\acknowledgements
We thank Takashi Nakamura and Masaru Shibata for fruitful discussions.
S.F. is supported by a Research Fellowship of Japan Society for the Promotion of Science (JSPS) for Young Scientists (No.26-1329).
This work has been supported by Grant-in-Aid for Scientific Research (No.23740160, No.24244028, and No.25103512) and by the HPCI Strategic Program of Japanese MEXT.

\bibliographystyle{apj}

\appendix
\section{Details of the reaction and cooling rates}\label{app:A}
In this appendix we summarize the cooling and reaction rates used to construct the neutrino-driven wind solutions.
The cooling rate due to electron and positron capture by free nucleons is given according to FFN85 as
\begin{equation}
q_{eN} = m_e \frac{\ln 2}{(ft)} \left( n_nJ_e|_{\rm pc}+n_pJ_e|_{\rm ec}\right),
\end{equation}
where $n_p$ and $n_n$ are the number densities of protons and neutrons, respectively, and $(ft)$ is the $ft$-value of the transition between nucleons, given as
\begin{eqnarray}
(ft)&=& \ln 2 \cdot \frac{2\pi^3}{G_{\rm F}^2(g_V^2 + 3g_A^2)m_e^5}\nonumber\\
&\approx&1.0157\times 10^3\ {\rm s}
\end{eqnarray}
where $G_{\rm F}$ is the Fermi coupling constant and $g_V= 1$ and $g_A=1.275$ are the vector-  and axial-vector-type coupling strength, respectively \citep[the values of them are taken from][]{2012PhRvC..86e5501P}.
$J_e$ is given using phase-space factor $J$ as
\begin{eqnarray}
J_e &=& \left(\frac{k_{\rm B}T}{m_e}\right)^6\int_{\eta^L_e}^\infty d\eta_e\ \eta_e^2(\eta_e + \zeta_n)^3\frac{1}{\exp(\eta_e - \eta_e^F)+1}\left[1-\frac{1}{\exp(\eta_e + \zeta_n - \eta_\nu^F)+1}\right]\nonumber\\
&=&\left(\frac{k_{\rm B}T}{m_e}\right)^6 \frac{1}{1-\exp(\eta_\nu^F - \zeta_n - \eta_e^F)}\ J.
\end{eqnarray}
The phase-space factor $J$ is given as
\begin{eqnarray}
J &=&\ F_5(\eta_e^F - \eta_e^L)-F_5(\eta_\nu^F - \zeta_n - \eta_e^L)\nonumber\\
&&+\ [5\eta_e^L + 3\zeta_n][F_4(\eta_e^F - \eta_e^L)-F_4(\eta_\nu^F - \zeta_n - \eta_e^L)]\nonumber\\
&&+\ [10(\eta_e^L)^2 + 12\eta_e^L\zeta_n + 3(\zeta_n)^2][F_3(\eta_e^F - \eta_e^L)-F_3(\eta_\nu^F - \zeta_n - \eta_e^L)]\nonumber\\
&&+\ [10(\eta_e^L)^3 + 18(\eta_e^L)^2\zeta_n + 9\eta_e^L(\zeta_n)^2 + (\zeta_n)^3][F_2(\eta_e^F - \eta_e^L)-F_2(\eta_\nu^F - \zeta_n - \eta_e^L)]\nonumber\\
&&+\ [5(\eta_e^L)^4 + 12(\eta_e^L)^3\zeta_n + 9(\eta_e^L)^2(\zeta_n)^2 + 2\eta_e^L(\zeta_n)^3][F_1(\eta_e^F - \eta_e^L)-F_1(\eta_\nu^F - \zeta_n - \eta_e^L)]\nonumber\\
&&+\ [(\eta_e^L)^5 + 3(\eta_e^L)^4\zeta_n + 3(\eta_e^L)^3(\zeta_n)^2 + (\eta_e^L)^2(\zeta_n)^3][F_0(\eta_e^F - \eta_e^L)-F_0(\eta_\nu^F - \zeta_n - \eta_e^L)],\nonumber\\
\end{eqnarray}
where $\eta_e^L$ is $\Delta/k_{\rm B}T$ in the case of electron capture by protons and is $m_e/k_{\rm B}T$ in the case of positron capture by neutrons, and $F_k(\eta)$ is the Fermi integral defined as
\begin{equation}
F_k(\eta) = \int_0^\infty dx\ x^k\frac{1}{\exp(x-\eta) + 1},
\end{equation}
which is evaluated using the approximated expressions in FFN85.

The cooling rate per unit mass due to electron-positron pair annihilation into a neutrino-antineutrino pair is given from \cite{1986ApJ...309..653C} as 
\begin{equation}
q_{e^+e^-} =\frac{1}{\rho}\bar{E}_{\rm loss}\Gamma_{e^+e^-},\label{eq:qdotee}
\end{equation}
where $\bar{E}_{\rm loss}$ is the average energy of the neutrino-antineutrino pair emitted per reaction and given as
\begin{equation}
\bar{E}_{\rm loss} = (k_{\rm B}T)\left[\frac{F_4(\eta_e^F)}{F_3(\eta_e^F)} + \frac{F_4(-\eta_e^F)}{F_3(-\eta_e^F)}\right],
\end{equation}
and $\Gamma_{e^+e^-}$ is the annihilation rate of the electron-positron pair per unit volume and given as
\begin{equation}
\Gamma_{e^+e^-}=\frac{G_{\rm F}^2 (D_e + D_\mu + D_\tau) }{9\pi} U_{e^-} U_{e^+}B_{e^+e^-}.
\end{equation}
$U_{e^\mp}$ is the energy density of electrons and positrons given as
\begin{equation}
U_{e^\mp}= \frac{(k_{\rm B}T)^4}{\pi^2} F_3(\pm \eta_e^F),
\end{equation}
and $B_{e^+e^-}$ is the final state blocking factor given using the average energy of emitted neutrino $\bar{E}_\nu=\bar{E}_{\rm loss}/2$ as
\begin{equation}
B_{e^+e^-}=\left[1+\exp\left( \eta^F_\nu - \frac{\bar{E}_\nu}{k_{\rm B}T} \right)\right]^{-1}\left[1+\exp\left( \eta^F_{\bar{\nu}} - \frac{\bar{E}_\nu}{k_{\rm B}T} \right)\right]^{-1}.
\end{equation}
This rate is derived using the approximation that the distributions of electrons and positrons are in thermal equilibrium (i.e., the chemical potential of positron is $-\mu_e$).
Equation (\ref{eq:qdotee}) can be written as
\begin{equation}
q_{e^+e^-} =\frac{1}{\rho}\frac{G_{\rm F}^2 (D_e + D_\mu + D_\tau)}{9\pi^5} (k_{\rm B}T)^9K.
\end{equation}
where the phase-space factor $K$ is given as
\begin{equation}
K=[F_4(\eta_e^F)F_3(-\eta_e^F) + F_4(-\eta_e^F)F_3(\eta_e^F)]B_{e^+e^-}.
\end{equation}

The reaction rate of electron and positron capture by free nucleons $\lambda_{\rm ec},\ \lambda_{\rm pc}$ is given from FFN85 as
\begin{equation}
\lambda=\frac{\ln 2}{(ft)}I_e,
\end{equation}
where $I_e$ is given using the phase-space factor of this reaction $I$ as 
\begin{eqnarray}
I_e &=& \left(\frac{k_{\rm B}T}{m_e}\right)^5\int_{\eta^L_e}^\infty d\eta_e\ \eta_e^2(\eta_e + \zeta_n)^2\frac{1}{\exp(\eta_e - \eta_e^F)+1}\left[1-\frac{1}{\exp(\eta_e + \zeta_n - \eta_\nu^F)+1}\right]\nonumber\\
&=&\left(\frac{k_{\rm B}T}{m_e}\right)^5 \frac{1}{1-\exp(\eta_\nu^F - \zeta_n - \eta_e^F)}\ I,
\end{eqnarray}
and $I$ becomes
\begin{eqnarray}
I &=& F_4(\eta_e^F - \eta_e^L) - F_4(\eta_\nu^F - \zeta_n - \eta_e^L)\nonumber\\
&&+\ (2\zeta_n + 4\eta_e^L)[F_3(\eta_e^F - \eta_e^L)-F_3(\eta_\nu^F - \zeta_n - \eta_e^L)] \nonumber\\
&&+\ [6(\eta_e^L)^2 + 6\zeta_n\eta_e^L + \zeta_n^2][F_2(\eta_e^F - \eta_e^L)-F_2(\eta_\nu^F - \zeta_n - \eta_e^L)]\nonumber\\
&&+\ [4(\eta_e^L)^3 + 6(\eta_e^L)^2\zeta_n + 2\eta_e^L(\zeta_n)^2][F_1(\eta_e^F - \eta_e^L)-F_1(\eta_\nu^F - \zeta_n - \eta_e^L)] \nonumber\\
&&+\ [(\eta_e^L)^4 + 2\zeta_n(\eta_e^L)^3 + (\zeta_n)^2(\eta_e^L)^2][F_0(\eta_e^F - \eta_e^L)-F_0(\eta_\nu^F - \zeta_n - \eta_e^L)].
\end{eqnarray}

\section{The treatment of the wind termination shock}\label{app:B}
In this appendix we describe the treatment of the wind termination shock.
Using the adiabatic sound speed $c_s^2 = (\partial P/\partial \rho)_s$ and the adiabatic gamma $\gamma = c_s^2/(P/\rho)$, the specific enthalpy (including the rest mass) $h$ and the pressure $P$ can be expressed as
\begin{eqnarray}
h = 1 + \frac{c_s^2}{\gamma-1},\ P = \frac{\rho c_s^2}{\gamma}.
\end{eqnarray}
We define the ratios $x$ and $y$ as
\begin{eqnarray}
x = \frac{\rho_{\rm s}}{\rho_{\rm w}} = \frac{u_{\rm w}}{u_{\rm s}},\ y = \frac{P_{\rm s}}{P_{\rm w}},
\end{eqnarray}
then Equations (\ref{eq:jump2}) and (\ref{eq:jump3}) become
\begin{eqnarray}
y &=& \frac{1+\gamma_{\rm w} M_{\rm w}^2 (1-1/x) +\gamma_{\rm w}u_{\rm w}^2/(1-\gamma_{\rm w})}{1+(1/x^2)\gamma_{\rm s}u_{\rm w}^2/(\gamma_{\rm s}-1)},\\
0&=&\left(x + y \frac{\gamma_{\rm s}}{\gamma_{\rm w}}\frac{c_{s,{\rm w}}^2}{\gamma_{\rm s}-1}\right)^2(x^2+u_{\rm w}^2) - x^4\left(1+\frac{c_{s,{\rm w}}^2}{\gamma_{\rm w}-1}\right)^2 (1+u_{\rm w}^2),
\end{eqnarray}
where $M=u/c_s$ is the Mach number of the fluid.
These equations are combined to give the equation of $x$ as
\begin{eqnarray}
\left[x + \frac{\gamma_{\rm s}}{\gamma_{\rm w}}\frac{c_{s,{\rm w}}^2}{\gamma_{\rm s}-1} \left(\gamma_{\rm w} M_{\rm w}^2 + 1 + \frac{\gamma_{\rm s}u_{\rm w}^2}{\gamma_{\rm s}-1}\right)\right]^2(x^2+u_{\rm w}^2) - \left(1+\frac{c_{s,{\rm w}}^2}{\gamma_{\rm w}-1}\right)^2 (1+u_{\rm w}^2)\left(x^2 + \frac{\gamma_{\rm s}u_{\rm w}^2}{\gamma_{\rm s}-1}\right)^2 = 0. \label{eq:solvex}
\end{eqnarray}
This equation has four solutions for $x$.
One of them gives the true post-shock quantities.
Another solution corresponds to the unshocked (no discontinuity) solution because this solution becomes $x=y=1$ if we assume $\gamma_{\rm s}=\gamma_{\rm w}$.
The other solutions are not real, so we do not consider them.
The solution related to the wind without the shock is close to unity, so we can obtain the post-shock solution by finding the real solution of Eq.(\ref{eq:solvex}) greater than another real solution, i.e., the unshocked one.

In this method, we assume the adiabatic gamma of the post-shock material $\gamma_{\rm s}$.
In general, it does not coincide with the adiabatic gamma evaluated from the obtained post-shock quantities.
In order to find the true post-shock quantities, we iteratively solve Equation (\ref{eq:solvex}) until the assumed $\gamma_{\rm s}$ coincides with the one evaluated from the post-shock quantities.
\end{document}